\documentclass[review]{elsarticle}

\usepackage{lineno,hyperref}
\usepackage{graphicx} % added package!
\usepackage{placeins} % added package!
\usepackage{graphicx,rotating,booktabs} % added package!
\usepackage{enumitem} % added package!
\usepackage{multirow} % added package!
\usepackage{subcaption}
\modulolinenumbers[5]

\journal{Applied Radiation and Isotopes}

\biboptions{longnamesfirst,semicolon,sort&compress}
\begin{document}

\begin{frontmatter}

\title{Neutron activation and prompt gamma intensity in Ar/CO$_{2}$-filled neutron detectors at the European Spallation Source}

% \address{Radarweg 29, Amsterdam}
% \fntext[myfootnote]{Since 1880.}

\author[mymainaddress,mysecondaryaddress,mytertiaryaddress]{E.~Dian \corref{mycorrespondingauthor}}
\cortext[mycorrespondingauthor]{Corresponding author}
\ead{dian.eszter@energia.mta.hu}
\author[mysecondaryaddress]{K.~Kanaki}
\author[mysecondaryaddress,myquaternaryaddress]{R.~J.~Hall-Wilton}
\author[mymainaddress,mytertiaryaddress]{P.~Zagyvai}
\author[mytertiaryaddress]{Sz.~Czifrus}

\address[mymainaddress]{Hungarian Academy of Sciences, Centre for Energy Research, 1525 Budapest 114., P.O. Box 49., Hungary}
\address[mysecondaryaddress]{European Spallation Source ESS ERIC, P.O Box 176, SE-221 00 Lund, Sweden}
\address[mytertiaryaddress]{Budapest University of Technology and Economics, Institute of Nuclear Techniques, 1111 Budapest, M\H uegyetem rakpart 9.}
\address[myquaternaryaddress]{Mid-Sweden University, SE-851 70 Sundsvall, Sweden}

\begin{abstract}
  %% % New 80 word abstract
  Monte Carlo simulations using MCNP6.1 were performed to study the effect of neutron activation in Ar/CO$_{2}$ neutron detector counting gas. A general MCNP model was built and validated with simple analytical calculations. Simulations and calculations agree that only the $^{40}$Ar activation can have a considerable effect. It was shown that neither the prompt gamma intensity from the $^{40}$Ar neutron capture nor the produced $^{41}$Ar activity have an impact in terms of gamma dose rate around the detector and background level.
%% Monte Carlo simulations using MCNP have been performed in order to study the effect of neutron activation in Ar/CO$_{2}$ neutron detector counting gas, from the perspective of decay gamma, as well as prompt gamma production. A general model for neutron activation has been built in MCNP6.1. Simple analytical calculations were also done to validate the full scale MCNP6.1 model. It has been shown that in accordance with our expectation, only the $^{40}$Ar activation could have a considerable effect in terms of radiation background. However, both the simulations and the calculations agree that either the prompt gamma intensity coming from the $^{40}$Ar neutron capture or the produced $^{41}$Ar activity %is still negligible.
%% are negligible in terms of gamma dose rate around detector and increasing background level.
\end{abstract}

\begin{keyword}
ESS \sep neutron detector \sep B4C \sep neutron activation \sep $^{41}$Ar \sep MCNP \sep Monte Carlo simulation
%%\MSC[2010] 00-01\sep  99-00
\end{keyword}

\end{frontmatter}

%\linenumbers

\section{Introduction}

Ar/CO$_{2}$ is a widely applied detector counting gas, with long history in radiation detection. Nowadays, the application of Ar/CO$_{2}$-filled detectors is extended in the field of neutron detection as well. However, the exposure of Ar/CO$_{2}$ counting gas to neutron radiation carries the risk of neutron activation. Therefore, detailed consideration of the effect and amount of neutron induced radiation in the Ar/CO$_{2}$ counting gas is a key issue, especially for large volume detectors.

In this paper methodology and results of detailed analytical calculations and Monte Carlo simulations of prompt and decay gamma production in boron-carbide-based neutron detectors filled with Ar/CO$_{2}$ counting gas are presented (see Appendix).

In Section~2 a detailed calculations method for prompt gamma and activity production and signal-to-background ratio is introduced, as well as a model built in MCNP6.1 for the same purposes. The collected bibliographical data (cross section, decay constant) and the cross section libraries used for MCNP6.1 simulation are also presented.

In Section~3 the results of the analytical calculations and the simulation, their comparison and their detailed analysis are given.

In Section~4 the obtained results are concluded from the aspects of gamma emission during and after irradiation, radioactive waste production and emission, and the effect of self-induced gamma background on the measured signal.

\section{Context}
\paragraph{The European Spallation Source (ESS) has the goal to be the world's leading neutron source for the study of materials by the second quarter of this century~\cite{ess,tdr}} %State of the art large-scale material-testing instruments are going to be served by the brightest neutron source of the world, performing 5~MW power on target.
Large scale material-testing instruments, beyond the limits of the current state-of-the-art instruments are going to be served by the brightest neutron source in the world, delivering 5~MW power on target.
%However,
At the same time the $^3$He crisis instigates detector scientists to open a new frontier for potential $^3$He substitute technologies and adapt them to the requirements of the large scale instruments that used to be fulfilled by well% tried and
-tested $^3$He detectors. One of the most promising replacements is the $^{10}$B converter based gaseous detector technology, utilising an Ar/CO$_{2}$ counting gas. Ar/CO$_{2}$-filled detectors will be utilised among others for inelastic neutron spectrometers~\cite{VORprop,VOR2015,CSPECprop,TREXprop}, where on the one hand, very large detector volumes are foreseen, on the other hand, very low background radiation is required. Consequently, due to the high incoming neutron intensity and large detector volumes, the effects of neutron-induced reactions, especially neutron activation in the solid body or in the counting gas of the detector could scale up and become relevant, both %either
in terms of background radiation and %or
radiation safety.

\paragraph{Gaseous detectors}
The gaseous ionisation chamber is one of the most common radiation detectors. The ionisation chamber itself is a gas filled tank that contains two electrodes with DC voltage~\cite{bodis,knoll5_6}. The detection method is based on the collision between atoms of the filling gas and the photons or charged particles to detect, during which electrons and positively charged ions are produced. Due to the electric field between the electrodes, the electrons drift to the anode, inducing a measurable signal. However, this measurable signal is very low, therefore typically additional are wires included and higher voltage is applied in order to obtain a gain on the signal, while the signal is still proportional to the energy of the measured particle; these are the so-called proportional chambers~\cite{sauli1977}. These detectors can be used as neutron detectors if appropriate converters are applied that absorb the neutron while emitting detectable particles via a nuclear reaction. In the case of $^3$He- or $^{10}$BF$_{3}$-based detectors the converter is the counting gas itself, but solid converters could be used as well with conventional counting gases.

\paragraph{Thermal neutron detector development} %The objective of the
One leading development has been set on the Ar/CO$_2$ gas filled detectors with solid enriched boron-carbide ($^{10}$B$_4$C) neutron converters~\cite{hoglund2012}, detecting neutrons via the $^{10}$B(n,$\alpha$)$^7$Li reaction~\cite{%khaplanov2013b
  andersen_2012,piscitelli2014,stefanescu2013,G.Nowak2015}. %The limitation of the existing
With this
technology %is that
the optimal thickness of the %%$^{10}$B$_4$C
boron-carbide layer is typically 1~$\mu$m~\cite{piscitelli2013}, otherwise the emitted $\alpha$ particle is stopped inside the layer and remains undetected. %However
But, a thinner %%$^{10}$B$_4$C
boron-carbide layer means smaller neutron conversion efficiency. The idea behind the detector development at ESS is the multiplication of %%$^{10}$B$_4$C
boron-carbide neutron converter layers by using repetitive geometrical structures, in order to increase the neutron conversion efficiency and obtain a detection efficiency that is competitive with %the one
that
of the $^3$He detectors~\cite{%khaplanov2013b
  andersen_2012,piscitelli2014}.

\paragraph{Shielding issues in detector development}
The modern neutron instruments are being developed to reach high efficiency, but also higher performance, such as time or energy resolution to open new frontiers in experimentation. One of the most representative characteristics of these instruments is the signal-to-background ratio, %that
which is targeted in the optimisation process. While the traditional solutions for improving the signal-to-background ratio are based on increasing the source power and improving the transmittance of neutron guides, for modern instruments the background reduction via optimised shielding becomes equally relevant. For state-of-the-art instruments the cost of a background reducing shielding can be a major contribution in the total instrument budget~\cite{cherkashyna_2015}. In order to optimise the shielding not only for radiation safety purposes but in order to improve the signal-to-background ratio a detailed map of potential background sources is essential. While the components of the radiation background coming from the neutron source and the neutron guide system are well known, the effect of newly developed %%$^{10}$B
boron-carbide converter based gaseous detectors still has to be examined, especially the background radiation and potential self-radiation coming from the neutron activation of the solid detector components and of the detector filling gas.

\paragraph{Argon activation}
The experience over the last decades showed that for facilities, e.g.~nuclear power plants, research reactors and research facilities with accelerator tunnels, there is a permanent activity emission during normal operation that mainly contains airborne radionuclides~\cite{hanaro,triga,BR1,BR12003,RA-3,srs43,ARCC,pwr}. For most of these facilities $^{41}$Ar is one of the major contributors %in
to the radiation release. $^{41}$Ar is produced via thermal neutron capture from the naturally occurring $^{40}$Ar, %that
which is the main isotope of natural argon with 99.3~\% abundance~\cite{toi}. At most %of the
facilities $^{41}$Ar is produced from the irradiation of the natural argon content of air. In air-cooled and water-cooled reactors $^{40}$Ar is exposed in the reactor core as part of the coolant; in the latter case it is coming from the air dissolved in the primary cooling water. 
Air containing argon is also present in the narrow %but existing
gap between the reactor vessel and the biological shielding. 
The produced $^{41}$Ar mixes with the air of the reactor hall and
is
removed by the ventilation system. In other facilities $^{41}$Ar is produced in the accelerator tunnel. In all cases, within the radiation safety plan of the facility the $^{41}$Ar release is taken into account~\cite{euratom} and well estimated either via simple analytical calculations or Monte Carlo simulations. The average yearly $^{41}$Ar release of these facilities can reach a few thousand GBq.

%Although 
For the ESS the $^{41}$Ar release coming from the accelerator and the spallation target is already calculated~\cite{radsafety,ene2011,andersson},
but in addition
the exposure of the large volume of Ar/CO$_2$ %that
contained in the neutron detectors should also be considered. Due to the 70-90~\% argon content of the counting gas and the fact that most instruments operate with thermal or cold neutron flux that leads to a higher average reaction rate, the $^{41}$Ar production in the detectors could be commensurate with the other sources. 
For all the above mentioned reasons, argon activation is an issue to consider at ESS both in terms of activity release and in terms of occupational exposure in the measurement hall.

\section{Applied methods}  
\paragraph{Analytical calculation for neutron activation}
%%The neutron capture with (n,$\gamma$) reaction gives the basics of two long-used and reliable analytical techniques, the neutron activation analysis (NAA) and the prompt gamma activation analysis (PGAA). Consequently, detailed measured and simulated data are available for neutron activation calculation.
Neutron activation occurs during the (n,$\gamma$) reaction where a neutron is captured by a target nucleus. The capture itself is usually followed by an instant photon emission; these are the so called prompt photons. The energies of the emitted prompt photons are specific for the target nucleus. After capturing the neutron, in most cases the nucleus gets excited, and becomes radioactive; this is the process of neutron activation, and the new radionuclide will suffer decay with its natural half-life. Due to their higher number of neutrons, the activated radionuclei mostly undergo $\rm \beta^-$~decay, accompanied by a well-measurable decay gamma radiation. The gamma energies are specific for the radionucleus. These two phenomena form the basics of two long-used and reliable analytical techniques, the neutron activation analysis (NAA~\cite{RadChem2003,Simonits1975,Simonits1989}) and the prompt gamma activation analysis (PGAA~\cite{molnar2004}). Consequently, detailed measured and simulated data are available for neutron activation calculation.

For shielding and radiation safety purposes the produced activity concentration ($a\rm~[Bq/cm^{3}]$) and the prompt photon intensity have to be calculated that are depending on the number of activated nuclei ($ N^{*}\rm~[1/cm^{3}]$). The production of radionuclides (\textit{reaction rate}) depends on the number of target nuclei ($ N_{0}\rm~[1/cm^{3}]$) for each relevant isotope, the irradiating neutron flux ($\Phi\rm~[n/cm^{3}/s]$) and the (n,$\gamma$) reaction cross section ($\sigma\rm~[cm^2]$) at the irradiating neutron energies, while the loss of radionuclides is determined by their decay constants ($\lambda\rm~[1/s]$). A basic assumption is that the number of target nuclei can be treated as constant if the loss of target nuclei during the whole irradiation %is not more than
does not exceed 0.1~\%. This condition is generally fulfilled, like in the cases examined in this study, therefore the rate of change of the number of activated nuclei is given by Equation~\ref{eq:n*}.

\begin{equation} \label{eq:n*}
  \frac{dN^{*}}{dt} =  N_{0}\cdot \Phi \cdot \sigma - \lambda \cdot N^{*} \\
\end{equation}

With the same conditions, the activity concentration after a certain time of irradiation ($t_{irr} [s]$) can be calculated with Equation~\ref{eq:atirr}.
\begin{equation}\label{eq:atirr}
  a \left(t_{irr}\right) = N_{0} \cdot \Phi \cdot \sigma \cdot \left(1- e^{\lambda t_{irr}}\right)
\end{equation}

In this study, as the activation calculation is based on Equation~\ref{eq:atirr}, the activity yield of the naturally present radionuclides (e.g.~cosmogenic~$^{14}$C in CO$_2$) is ignored due to the very low abundance of these nuclides. The activity yield of the secondary activation products, the products of multiple independent neutron captures on the same target nucleus, are ignored as well, because of the low probability of the multiple interaction.

The prompt gamma intensity ($I\rm~[1/s/cm^3$]) coming from the neutron capture can be calculated similarly to the (n,$\gamma$) reaction rate. In this case a prompt gamma line ($i$) specific cross section ($\sigma_{pg,i}$) has to be used~\cite{iaeapgaa}, that is proportional to the (n,$\gamma$) cross section, the natural abundance of the target isotope in the target element, and the weight of the specific gamma energy with respect to the total number of gamma lines. For this reason in Equation~\ref{eq:pg} the number of target nuclei corresponds to the element ($ N'_{0}\rm~[1/cm^{3}]$), not the isotope ($ N_{0}~\rm[1/cm^{3}]$). 

\begin{equation} \label{eq:pg}
  I_{i} =  N'_{0}\cdot \Phi \cdot \sigma_{pg,i}  \\
\end{equation}

In the current study, activity concentration, prompt gamma intensity and the respective prompt gamma spectrum have been calculated for each isotope in the natural composition~\cite{toi} of an 80/20~volume~ratio Ar/CO$_2$ counting gas at room temperature and 1~bar pressure and in an aluminium alloy %as
used for the detector frame. Alloy Al5754~\cite{al5754} has been chosen as a typical alloy used in nuclear science for mechanical structures.
Activity concentration and prompt gamma intensity calculations have been done for several monoenergetic neutron beams in the range of 0.6--10~\AA \ (227.23--0.82~meV). Since for isotopes of interest the energy dependence of the (n,$\gamma$) cross section is in the 1/v (velocity) region~\cite{iaeatd1285,iaeands}, the cross sections for each relevant energy %%could
have been easily extrapolated from the thermal (1.8~\AA) neutron capture cross sections listed in Table~\ref{tab:siglambda}.

The irradiating neutron flux has been approximated with $\rm 10^{4}~n/cm^2/s$. This %number
value has been determined for a worst case scenario based on the following assumptions: the planned instruments are going to have various neutron fluxes at the sample position, and the highest occurring flux can be conservatively estimated to $\rm 10^{10}~n/cm^2/s$ \cite{c.hoglund2015b}. The neutron fraction scattered from the sample is in the range of 1-10~\%. Calculating with 10~\%, the approximation remains conservative. A realistic sample surface is 1~cm$^2$, %that is
reducing the scattered flux to  $\rm 10^{9}$ n/s. The sample-detector distance also varies  among the instruments, so the smallest realistic distance of 100~cm was used for a conservative approximation. Therefore the neutron yield has to be normalised to a $\rm 10^5~cm^2$ surface area at this sample-detector distance. According to these calculations, $\rm 10^4~n/cm^2/s$ is a conservative estimation for the neutron flux the detector is exposed to. This simple approach allows that the result can be scaled to alternate input conditions, i.e. a higher neutron flux or detector geometry.

\paragraph{MCNP simulation for neutron activation}
Monte Carlo simulations have been performed in order to determine the expected activity concentration and prompt gamma intensity in the counting gas
and the aluminium frame
of boron-carbide-based %large area
neutron detectors.% filled with Ar/CO$_{2}$. 

\begin{figure}[ht!]
\centering
\includegraphics[width=120mm]{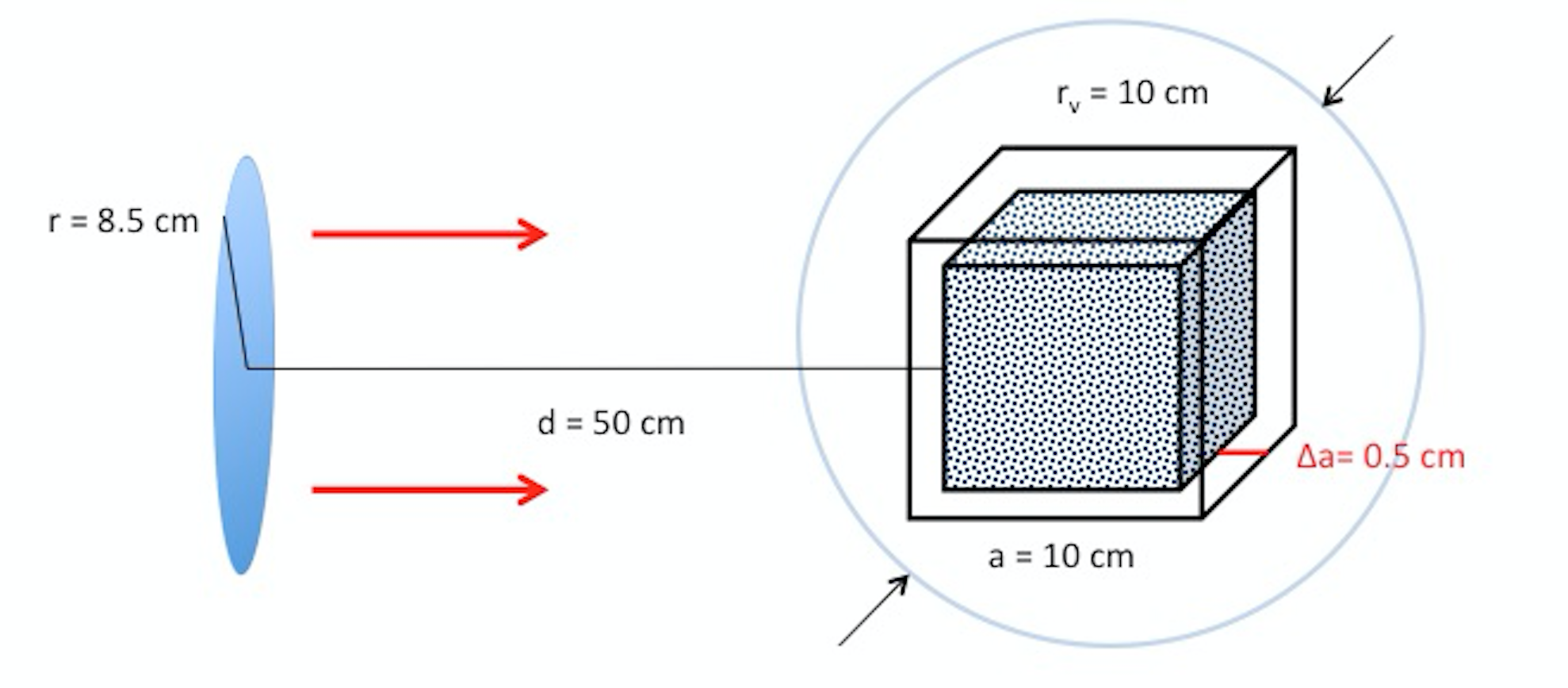}
\caption{Neutron irradiation geometry used in MCNP6 simulation. %Detector frame was approximated as cube with 5~mm thick walls around the 10~x~10~x~10~cm$^3$ cubic gas volume, irradiated with monoenergetic neutron beam from a monodirectional disk source of 8.5~cm radius and 50~cm distance from the target gas volume in vacuum. A virtual sphere has been defined around the target gas volume with 10~cm radius for simplifying prompt photon counting.
  \label{geom_al} }
\end{figure}

The MCNP6.1~\cite{mcnpman} version has been used for the simulations. The detector gas volume has been approximated %within the simulations
as a generic 10~cm~x~10~cm x~10~cm cube, surrounded by a 5~mm thick aluminium box made of Al5754 alloy, representing the detector frame, as it is described in Figure~\ref{geom_al}. %Al5754 alloy \cite{al5754} has been choosen as a typical detector material.
In order to avoid interference with the prompt photon emission of the Ar/CO$_2$, the counting gas was replaced with vacuum while observing the activation on the aluminium frame.  
The detector geometry has been irradiated with a monoenergetic neutron beam from a monodirectional disk source of 8.5~cm radius at 50~cm distance from the surface of the target volume. A virtual sphere has been defined around the target gas volume with 10~cm radius for simplifying prompt photon counting. Both the activity concentration and the prompt gamma intensity determined with MCNP6.1 simulations have been scaled to a $\rm 10^4~n/cm^2/s$ irradiating neutron flux.

Different runs have been prepared for each element in the gas mixture
and the Al5754 alloy 
to determine the prompt gamma spectrum and total intensity. The prompt photon spectrum has been determined for each element with the following method: a virtual sphere %with $\rm r = 10~cm$ radius
has been defined around the cubic target volume. Since the target volume was located in vacuum, all the prompt photons produced in a neutron activation reaction have to cross this virtual surface. Within MCNP, the particle current integrated over a surface, can be easily determined (F1~tally \cite{mcnpman}). Knowing the volume of the target, the prompt photon intensity can be calculated for the simulated neutron flux ($\Phi_{MCNP}$, [flux/source particle]). After the $\Phi_{MCNP}$ average neutron flux in the target volume has been determined (F4~tally \cite{mcnpman}), the prompt photon intensity can be scaled for any desired neutron flux, $\rm 10^{4}~n/cm^2/s$ in this case. With this method the self-absorption of the target gas volume can be considered to be negligible.

%%\newpage
The activity concentration is not given directly by the simulation, but it can be calculated from the $R_{MCNP}$~reaction rate (reaction/source particle) and the $\Phi_{MCNP}$~flux. The $R_{MCNP}$ is calculated in MCNP in the following way: first the track length density of neutrons has to be determined in the target volume (F4~tally~\cite{mcnpman}), and then this value has to be multiplied with the reaction cross section of the specific reaction of interest, through the entire spectrum, taking into account the number of target nuclei of the irradiated material (FM tally multiplication card~\cite{mcnpman}). In the current simulations each isotope has been defined as a different material, with their real partial atomic density ([atom/barn/cm]) in the counting gas or in the aluminium alloy for the (n,$\gamma$) reaction (ENDF reaction 102). As the reaction rate given by the MCNP simulation is the saturated reaction rate for the $\Phi_{MCNP}$~flux, and contains all the geometrical and material conditions of the irradiation, the time-dependent activity concentration for any $\Phi$~flux can be calculated with Equation~\ref{eq:amcnp}.

\begin{equation}\label{eq:amcnp}
  a \left(t_{irr}\right) = R_{MCNP} \cdot \frac{\Phi}{\Phi_{MCNP}} \cdot \left(1- e^{\lambda t_{irr}}\right)
 %% flux/flux_mcnp * RR[i] * (1 - m.exp((-1) * element[i]['hl_s'] * j) )
\end{equation}

%The cardinal issue
%%One of the main issues of any MCNP simulation is the usage of correct cross section libraries.
In order to determine the above mentioned quantities, the cross section libraries have to be chosen carefully for the simulation.
Within the current study different libraries have been used to simulate the prompt gamma production and the reaction rates. Several databases have been tested, but only a few of them contain data on photon production for the isotopes of interest.
Tables~\ref{tab:crosseclib}~and~\ref{tab:crosseclib_al}
present 
the combinations %are shown
that give the best agreement with the theoretical expectations, especially in terms of spectral distribution. These are the ENDF~\cite{iaeands}, TALYS~\cite{talys} and LANL~\cite{mcnplib} databases.

The MCNP6.1 simulation has been repeated for each naturally occurring isotope in the counting gas and the aluminium frame, and analytical calculations have been also prepared to validate the simulation, in order to obtain reliable and well-applicable data on the detector housing and counting gas activation and gamma emission both for shielding and for radiation protection purposes.

%%\newpage

In order to demonstrate the effect of gamma radiation on the measured signal, the signal-to-background has been calculated for a typical and realistic detector geometry. A generic boron-carbide based detector can be represented by a 5-20~mm thick gas volume surrounded by a few millimetre thin aluminium box, carrying the few micrometers thick boron-carbide converter layer(s). The gas volume is determined by the typical distance needed for the energy deposition. In a realistic application, a larger gas volume used to be used for efficiency purposes, built up from the above mentioned subvolumes.
%A single grid of the Multi-Grid detector have been choosen~\cite{khaplanov2013b}, build with 16 double layer of 1~$\mu$m enriched $^{10}$B$_4$C as converter, dividing the $\rm V_{gas}$ gas volume into 4~x~16 gas cells, each with the size of 2~cm~x~2~cm~x~1~cm, as it is shown in Figure~\ref{sgrid}. This geometry leading to $\rm V_{gas} = 256 \ cm^3$ gas volume as source of gamma production, and $\rm A_{in} = 16 \ cm^2$ entrance surface for incident neutrons.
As a representative example a $V_{gas} = 256 \ cm^3$ counting gas volume has been chosen as the source of gamma production, with an  $A_{in} = 16 \ cm^2$ entrance surface for incident neutrons, divided into 20~mm thick subvolumes by 16 layers of 2~$\mu$m thin enriched %$^{10}$B$_4$C.
boron-carbide. \label{sec:SNR}

In this study the gamma efficiency has been approximated with $\rm 10^{-7}$ for the entire gamma energy range~\cite{khaplanov2014, khaplanov2013a} due to its relatively low energy-dependence, whereas the neutron efficiency has been calculated for all the mentioned energies on the basis of~\cite{andersen_2012}%~\cite{khaplanov2013b}
, resulting in a neutron efficiency varying between 0.4-0.72 within the given energy range. Therefore the measured signal and the signal of the gamma background were calculated as in Equations~\ref{eq:S_n}-\ref{eq:S_g}, where $\eta_{i}$ is the detection efficiency for the particle type $i$, $\Phi$ is the incident neutron flux and $I_{photon}$ is the produced photon production in a unit gas volume. Signal-to-(gamma-)background ratio has been calculated as $S_n/S{\gamma}$.

\begin{equation}\label{eq:S_n}
  S_{n} = A_{in} \cdot \Phi \cdot \eta_{n}
\end{equation}

\begin{equation}\label{eq:S_g}
  S_{\gamma} = V_{gas} \cdot I_{photon} \cdot \eta_{\gamma}
\end{equation}

All calculations and simulations have been done for a $\rm 10^4~n/cm^2/s$ monoenergetic neutron irradiation for 0.6, 1, 1.8, 2, 4, 5 and 10~$\rm \AA$ neutron wavelengths. Activity concentration has been calculated for $t_{irr}\rm = 10^{6}~s$ irradiation time and $t_{cool}\rm = 10^{7}~s$ cooling time. 
This irradiation time roughly corresponds to typical lengths of operation cycles for spallation facilities.
%In some cases photon production has been normalised to incident neutron flux  meaning a $\rm \Phi = 1~n/cm^2/s$ neutron flux irradiating a 1~cm$^3$ volume.
Photon production has been normalised for a 1~cm$^3$ volume, irradiated with $\Phi = 1\rm~n/cm^2/s$ or $\Phi = 10^{4}\rm~n/cm^2/s$ neutron flux. 
Therefore here the photon production in a unit gas or aluminium volume irradiated with a unit flux is given as $\rm \frac{photon/cm^3/s}{n/cm^2/s}$.

The uncertainties of the simulated and the bibliographical data have been taken into account. The MCNP6.1 simulations had high enough statistics, that the uncertainties of the simulated results were comparable to the uncertainties of the measured/bibliographical qualities used for the analytical calculations. The uncertainty of the total prompt photon production for all elements were below 5~\% for the entire neutron energy range, while the uncertainties of the main prompt gamma lines were below 10~\% for all elements, and less than 5~\% for argon and the elements of the aluminium alloy.

For the anaytical calculations, %the following uncertainities have been used:
the error propagation takes into account the uncertainty of %the one of
the prompt gamma line specific cross section, given in the IAEA PGAA Database~\cite{iaeapgaa}, being below 5~\% for the main lines of all major isotopes, the $\sigma$ absorption cross section and the $\lambda$ decay constant (see Appendix). %%through the Absolute Error Propagation Rule.
The obtained uncertainties of the photon intensities are generally within the size of the marker, here the error bars have been omitted. They have also been omitted for some of the spectra for better visibility.

%%\newpage
\section{Results and discussion} 

%% REVISED!
%%\paragraph{Prompt gamma intensity in detector counting gas}
\subsection{Prompt gamma intensity in detector counting gas}

%%An analytical calculation (Equation~\ref{eq:pg}) based on detailed prompt gamma data from IAEA PGAA  Data-base~\cite{iaeapgaa} has been done in order to determine the total prompt photon production and its spectral distribution in Ar/CO$_{2}$ counting gas.
%%The same data have been obtained with Monte Carlo simulation using MCNP6.1.
 The total prompt photon production and its spectral distribution in Ar/CO$_{2}$ counting gas has been analytically calculated (Equation~\ref{eq:pg}) on the basis of detailed prompt gamma data from IAEA PGAA  Data-base~\cite{iaeapgaa}.
The same data have been obtained with Monte Carlo simulation using MCNP6.1.

Prompt photon production normalised to incident neutron flux has been calculated for all mentioned wavelengths. %By comparing results it has been
The comparison of the result has shown, that the simulated and calculated total prompt photon yields qualitatively agree for Ar, C, and O within 2~\%, 11~\% and 21~\%, respectively.
\FloatBarrier
\begin{figure}[ht!]
  \centering
  \includegraphics[width=\textwidth]{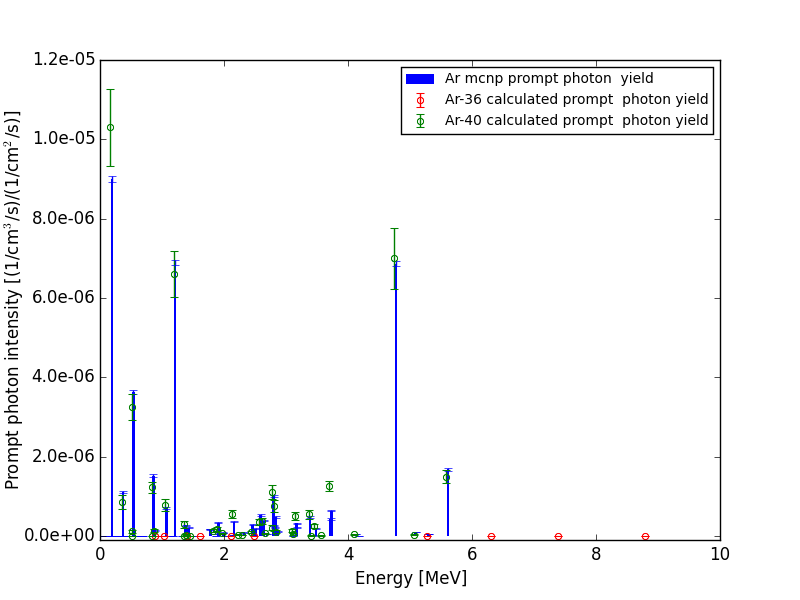}
  \caption{Prompt photon emission spectra from Ar in Ar/CO$_2$, irradiated with unit flux of 1.8~$\rm \AA$ neutrons. Results of analytical calculation %%REVISED!on the basis of
    with input data taken from IAEA PGAA Database~\cite{iaeapgaa} and MCNP6.1 simulation, as explained in text. (For interpretation of the references to color in this figure caption, the reader is referred to the web version of this article.) \label{Ar_all_P}}
\end{figure}

It has also been show that for these three elements proper cross section libraries can be found (see Table~\ref{tab:crosseclib}), the use of which in MCNP simulations produce prompt photon spectra that qualitatively agree with the calculated ones. As an example Figure~\ref{Ar_all_P} shows the simulated and calculated prompt photon spectra from argon in Ar/CO$_{2}$ for a 1.8~\AA, $\Phi\rm = 1~n/cm^2/s$ neutron flux, irradiating a 1~cm$^3$ volume.  Since numerous databases lack proper prompt photon data, this agreement is not trivial to achieve for all the elements. %It has been shown that
For these three elements MCNP simulations can effectively replace analytical calculations, which is especially valuable for more complex geometries. For all these reasons hereinafter only the MCNP6.1 simulated results are presented.
\FloatBarrier
\begin{figure}[ht!]
\centering
\includegraphics[width=\textwidth]{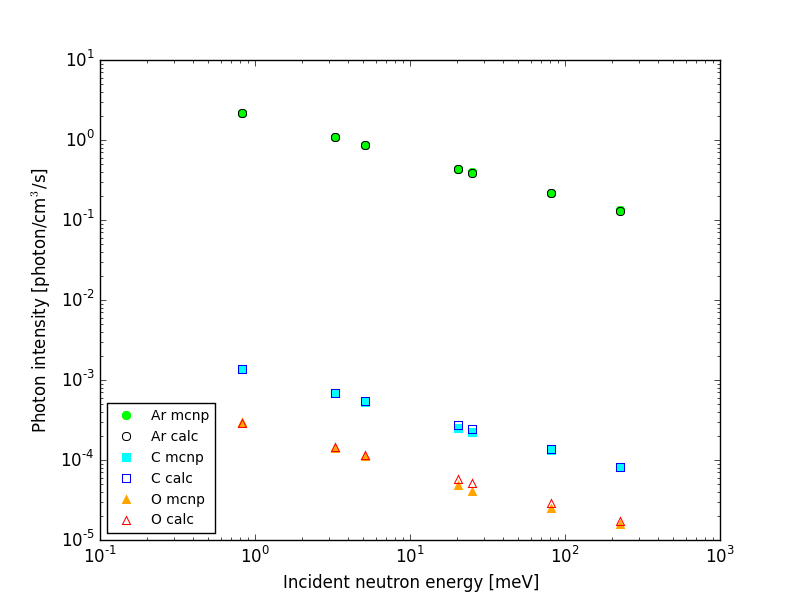}
\caption{Elemental distribution of total prompt photon intensity in Ar/CO$_2$ counting gas irradiated with $\rm 10^{4}~n/cm^{2}/s$ neutron flux. Results of MCNP6.1 simulation and analytical calculations %%REVISED! based on
  with input data taken from IAEA PGAA Database~\cite{iaeapgaa}, as explained in text. (For interpretation of the references to color in this figure caption, the reader is referred to the web version of this article.) \label{arco2_p}}
\end{figure}

 In Figure~\ref{arco2_p} it is also shown, that the prompt photon emission is dominated by argon, as expected due to the very small capture cross section of the oxygen and the carbon; the argon total prompt photon yield is 3 orders of magnitude higher than the highest of the rest. According to Figure~\ref{Ar_all_P}, within the argon prompt gamma spectrum, there are 3 main gamma lines that are responsible for the majority of the emission; %the ones with 167.30~$\pm$~20~keV, 1186.8~$\pm$~3~keV and 4745.3~$\pm$~8~keV energies.
the ones at 167~$\pm$~20~keV, 1187~$\pm$~3~keV and 4745~$\pm$~8~keV. %energies.

%%2016.11.17.

%%\newpage
%% REVISED!
%%\paragraph{Activity concentration and decay gammas in detector counting gas}
\subsection{Activity concentration and decay gammas in detector counting gas}
%%An analytical calculation has been done in order to determine the induced activity in the irradiated Ar/CO$_{2}$ gas volume, as well as the photon yield coming from the activated radionuclei. The calculation was based on the bibliographical thermal (25.30~meV) neutron capture cross sections and the half-lives of the isotopes in the counting gas (see Table~\ref{tab:siglambda}). A similar calculation has been done on the bases of reaction rates determined with MCNP simulation for each isotope of the counting gas. Activity concentrations obtained from calculation and MCNP6.1 simulation agree within the margin of error, therefore only the MCNP simulations are presented.
The induced activity in the irradiated Ar/CO$_{2}$ gas volume, as well as the photon yield coming from the activated radionuclei has been determined via analytical calculation, based on the bibliographical thermal (25.30~meV) neutron capture cross sections and the half-lives of the isotopes in the counting gas (see Table~\ref{tab:siglambda}). A similar calculation has been prepared on the bases of reaction rates determined with MCNP simulations for each isotope of the counting gas. Activity concentrations obtained from the calculation and the MCNP6.1 simulation agree within the margin of error, therefore only the MCNP simulations are presented.

%Total and isotopical activity concentrations have been calculated for a $\rm 10^6~s$ irradiation time (11~days), followed by a $\rm 10^7~s$ (0.3~year) cooling time, for a $\rm 10^4~n/cm^2/s$ monoenergetic irradiating neutron flux at 0.6, 1, 1.8, 2, 4, 5 and 10~$\rm \AA$ wavelengths.

\begin{figure}[ht!]
\centering
\includegraphics[width=\textwidth]{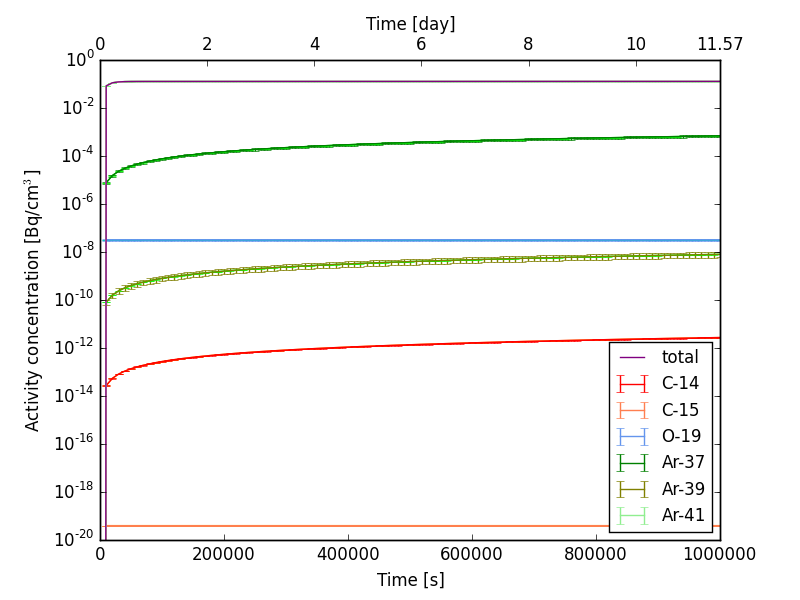}
\caption{Build-up of isotopic and total activity concentration [Bq/cm$^3$] in Ar/CO$_2$ during 10$^6$~s irradiation time. Results of MCNP6.1 simulation% and analytical calculations
  , as explained in text% \cite{mugh81} \cite{toi}
  . (For interpretation of the references to color in this figure caption, the reader is referred to the web version of this article.) \label{arco2_act}}
\end{figure}

As an example the build-up of activity during irradiation time for 1.8~$\rm \AA$ is given in Figure~\ref{arco2_act} for all the produced radionuclei.
%According to Figures~\ref{arco2_act}~and~\ref{arco2_decay}, the
\FloatBarrier

It can be stated, that the total activity of the irradiated counting gas practically equals the $^{41}$Ar activity (see Figure~\ref{arco2_act}), which is $\rm 1.28 \cdot 10^{-1} \ Bq/cm^3$ at the end of the irradiation time. This is 2~orders of magnitude higher than the activity of $^{37}$Ar, which is $\rm 6.90 \cdot 10^{-4} \ Bq/cm^3$, and 7~orders of magnitude higher than the activity of $^{38}$Ar ($\rm 7.99 \cdot 10^{-9} \ Bq/cm^{3}$) and $^{19}$O ($\rm 3.19 \cdot 10^{-8} \ Bq/cm^{3}$). The activity of carbon is negligible. 

\begin{figure}[ht!]
\centering
\includegraphics[width=\textwidth]{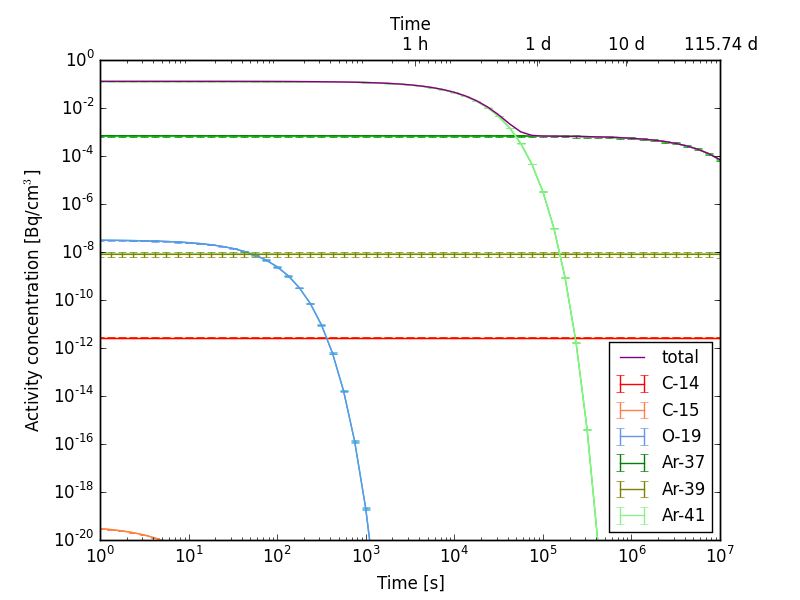}
\caption{Decrease of activity concentration [Bq/cm$^3$] in Ar/CO$_2$ from end of the 10$^6$~s irradiation period. Results of MCNP6.1 simulation% and analytical calculations
  , as explained in text% \cite{mugh81} \cite{toi}
  . (For interpretation of the references to color in this figure caption, the reader is referred to the web version of this article.) \label{arco2_decay}}
\end{figure}

The decrease of activity in the detector counting gas because of the natural radioactive decay is shown in Figure~\ref{arco2_decay}. %As expected,
After the end of the irradiation the main component of the total activity is the $^{41}$Ar, although it practically disappears after a day ($\rm 10^5$~s), due to its short 109.34~m half-life %of it and then the
with $^{37}$Ar becoming the dominant isotope. However, in terms of gamma emission, all the remaining isotopes, $^{37}$Ar, $^{39}$Ar and $^{14}$C are irrelevant, since they are pure beta-emitters. Therefore, %it can be stated that
with the above listed conditions there is only minimal gamma emission from the Ar/CO$_2$ counting gas after $\rm 10^5$~s cooling time. For the same reason, the $^{41}$Ar activity quickly saturates and %therefore
accordingly it can contribute to the gamma emission during the irradiation as well.

%%2016.11.17.
%Figures~\ref{Ar_D_calc}-\ref{O_D_calc} depict
Decay gamma emission of the activated radionuclei from a unit volume per second, with the activity reached by the end of the irradiation time have been calculated. It is shown that the decay gamma yield practically all comes from the activated argon; the emission of the 1293.587~keV $^{41}$Ar line is 8~orders of magnitudes higher than the yield of any other isotope. 

Comparing the prompt and the maximum decay gamma emission of all the isotopes, as it is shown in Table~\ref{tab:yield}, it is revealed that for the argon, the prompt photon production ($\rm 3.9 \cdot 10^{-1} \ \frac{photon/cm^3/s}{n/cm^2/s}$) and the saturated decay gamma production ($\rm 1.27 \cdot 10^{-1} \ \frac{photon/cm^3/s}{n/cm^2/s}$) are comparable. There is a factor of 3 difference, whereas for carbon and oxygen the decay gamma production is negligible comparing with the prompt gamma production.

\begin{sidewaystable}[htbp]
  %%\tiny
  %%\resizebox{\textwidth}{!}
  \centering
  \caption{Prompt and decay gamma emission from 80/20~V\% Ar/CO$_{2}$ at 1~bar pressure and from Al5754 aluminium alloy, irradiated with $\rm 10^{4} \ \frac{1}{cm^{2} \ s}$ monoenergetic neutron flux for $\rm 10^{6} \ s$ irradiation time.  Results of MCNP6.1 simulation. \label{tab:yield} }
  \label{tab:Iph_g}
  \resizebox{\textwidth}{!}{\begin{tabular}{ccccccccc}
      %%  \begin{tabular}{ccccccccc}
      \hline
      Element & Photon yield & \multicolumn{7}{c}{Neutron wavelength [\AA]} \\
      &  $\left[ \frac{1}{cm^{3} \ s} \right]$  & 0.6 & 1 & 1.8 & 2 & 4 & 5 & 10  \\
      %% Element & Photon yield & \multicolumn{21}{c}{Neutron wavelength [\AA]} \\
      %% &  $\left[ \frac{1}{cm^{3} \ s} \right]$  & & 0.6 & & & 1 & & & 1.8 & & & 2 & & & 4 & & & 5 & & & 10 &  \\
      \hline
      \multirow{2}{*}{Ar}
      & prompt & $\rm 1.32\phantom{00} \pm 0.04\phantom{00} \cdot 10^{-1\phantom{0}} $ & $\rm 2.15\phantom{00} \pm 0.05\phantom{00} \cdot 10^{-1\phantom{0}} $ & $\rm 3.96\phantom{00} \pm 0.08\phantom{00} \cdot 10^{-1\phantom{0}} $ & $\rm 4.37\phantom{00} \pm 0.09\phantom{00} \cdot 10^{-1\phantom{0}} $ & $\rm 8.64\phantom{00} \pm 0.14\phantom{00} \cdot 10^{-1\phantom{0}} $ & $\rm 1.080\phantom{0} \pm 0.016\phantom{0} \cdot 10^{0\phantom{-0}} $ & $\rm 2.150\phantom{0} \pm 0.025\phantom{0} \cdot 10^{0\phantom{-0}} $\\
      & decay  & $\rm 4.227\phantom{0} \pm 0.001\phantom{0} \cdot 10^{-2\phantom{0}} $ & $\rm 7.045\phantom{0} \pm 0.002\phantom{0} \cdot 10^{-2\phantom{0}} $ & $\rm 1.2667 \pm 0.0003 \cdot 10^{-1\phantom{0}} $ & $\rm 1.4090 \pm 0.0004 \cdot 10^{-1\phantom{0}} $ & $\rm 2.8179 \pm 0.0007 \cdot 10^{-1\phantom{0}} $ & $\rm 3.5224 \pm 0.0009 \cdot 10^{-1\phantom{0}} $ & $\rm 7.044\phantom{0} \pm 0.002\phantom{0} \cdot 10^{-1\phantom{0}} $\\
      \multirow{2}{*}{C}
      & prompt & $ \rm 8.1\phantom{000} \pm 1.4\phantom{000} \cdot 10^{-5\phantom{0}} $ & $\rm 1.33\phantom{00} \pm 0.18\phantom{00} \cdot 10^{-4\phantom{0}} $ & $\rm 2.21\phantom{00} \pm 0.23\phantom{00} \cdot 10^{-4\phantom{0}} $ & $\rm 2.51\phantom{00} \pm 0.25\phantom{00} \cdot 10^{-4\phantom{0}} $ & $\rm 5.33\phantom{00} \pm 0.36\phantom{00} \cdot 10^{-4\phantom{0}} $ & $\rm 6.9\phantom{000} \pm 0.4\phantom{000} \cdot 10^{-4\phantom{0}} $ & $ \rm 1.36\phantom{00} \pm 0.06\phantom{00} \cdot 10^{-3\phantom{0}} $\\
      & decay  & $ \rm 8.49\phantom{00} \pm 0.11\phantom{00} \cdot 10^{-21} $ & $\rm 1.44\phantom{00} \pm 0.02\phantom{00} \cdot 10^{-20} $ & $\rm 2.51\phantom{00} \pm 0.03\phantom{00} \cdot 10^{-20} $ & $\rm 2.79\phantom{00} \pm 0.04\phantom{00} \cdot 10^{-20} $ & $\rm 5.56\phantom{00} \pm 0.07\phantom{00} \cdot 10^{-20} $ & $\rm 6.94\phantom{00} \pm 0.09\phantom{00} \cdot 10^{-20} $ & $\rm 1.39\phantom{00} \pm 0.02\phantom{00} \cdot 10^{-19} $\\
      \multirow{2}{*}{O}
      & prompt & $ \rm 1.58\phantom{00} \pm 0.43\phantom{00} \cdot 10^{-5\phantom{0}} $ & $\rm 2.51\phantom{00} \pm 0.55\phantom{00} \cdot 10^{-5\phantom{0}} $ & $\rm 4.1\phantom{000} \pm 0.7\phantom{000} \cdot 10^{-5\phantom{0}} $ & $\rm 4.81\phantom{00} \pm 0.77\phantom{00} \cdot 10^{-5\phantom{0}} $ & $\rm 1.12\phantom{00} \pm 0.12\phantom{00} \cdot 10^{-4\phantom{0}} $ & $\rm 1.43\phantom{00} \pm 0.14\phantom{00} \cdot 10^{-4\phantom{0}} $ & $\rm 2.96\phantom{00} \pm 0.19\phantom{00} \cdot 10^{-4\phantom{0}} $ \\
      & decay  & $ \rm 1.619\phantom{0} \pm 0.035\phantom{0} \cdot 10^{-8\phantom{0}} $ & $\rm 2.70\phantom{00} \pm 0.06\phantom{00} \cdot 10^{-8\phantom{0}} $ & $\rm 4.8\phantom{000} \pm 0.1\phantom{000} \cdot 10^{-8\phantom{0}} $ & $\rm 5.40\phantom{00} \pm 0.12\phantom{00} \cdot 10^{-8\phantom{0}} $ & $\rm 1.08\phantom{00} \pm 0.02\phantom{00} \cdot 10^{-7\phantom{0}} $ & $\rm 1.35\phantom{00} \pm 0.03\phantom{00} \cdot 10^{-7\phantom{0}} $ & $\rm 2.69\phantom{00} \pm 0.06\phantom{00} \cdot 10^{-7\phantom{0}} $\\
      \hline
      \multirow{2}{*}{Al}
      & prompt & $\rm 8.27\phantom{00} \pm 0.11\phantom{00} \cdot 10^{1\phantom{-0}} $ & $\rm 1.379\phantom{0} \pm 0.015\phantom{0} \cdot 10^{2\phantom{-0}} $ & $\rm 2.47\phantom{00} \pm 0.02\phantom{00} \cdot 10^{2\phantom{-0}} $ & $\rm 2.75\phantom{00} \pm 0.02\phantom{00} \cdot 10^{2\phantom{-0}} $ & $\rm 5.44\phantom{00} \pm 0.03\phantom{00} \cdot 10^{2\phantom{-0}} $ & $\rm 6.76\phantom{00} \pm 0.03\phantom{00} \cdot 10^{2\phantom{-0}} $ & $\rm 1.300\phantom{0} \pm 0.005\phantom{0} \cdot 10^{3\phantom{-0}} $\\
      & decay  & $\rm 4.4419 \pm 0.0018 \cdot 10^{1\phantom{-0}} $ & $\rm 7.401\phantom{0} \pm 0.003\phantom{0} \cdot 10^{1\phantom{-0}} $ & $\rm 1.3288 \pm 0.0005 \cdot 10^{2\phantom{-0}} $ & $\rm 1.4773 \pm 0.0006 \cdot 10^{2\phantom{-0}} $ & $\rm 2.929\phantom{0} \pm 0.001\phantom{0} \cdot 10^{2\phantom{-0}} $ & $\rm 3.6373 \pm 0.0015 \cdot 10^{2\phantom{-0}} $ & $\rm 6.9981 \pm 0.0028 \cdot 10^{2\phantom{-0}} $\\
      \multirow{2}{*}{Cr}
      & prompt & $\rm 2.0\phantom{000} \pm 0.1\phantom{000} \cdot 10^{0\phantom{-0}} $ & $\rm 3.35\phantom{00} \pm 0.14\phantom{00} \cdot 10^{0\phantom{-0}} $ & $\rm 6.0\phantom{000} \pm 0.2\phantom{000} \cdot 10^{0\phantom{-0}} $ & $\rm 6.7\phantom{000} \pm 0.2\phantom{000} \cdot 10^{0\phantom{-0}} $ & $\rm 1.34\phantom{00} \pm 0.03\phantom{00} \cdot 10^{1\phantom{-0}} $ & $\rm 1.680\phantom{0} \pm 0.036\phantom{0} \cdot 10^{1\phantom{-0}} $ & $\rm 3.35\phantom{00} \pm 0.05\phantom{00} \cdot 10^{1\phantom{-0}} $\\
      & decay  & $\rm 5.4774 \pm 0.0026 \cdot 10^{-3\phantom{0}} $ & $\rm 9.131\phantom{0} \pm 0.004\phantom{0} \cdot 10^{-3\phantom{0}} $ & $\rm 1.6418 \pm 0.0008 \cdot 10^{-2\phantom{0}} $ & $\rm 1.8263 \pm 0.0009 \cdot 10^{-2\phantom{0}} $ & $\rm 3.653\phantom{0} \pm 0.002\phantom{0} \cdot 10^{-2\phantom{0}} $ & $\rm 4.566\phantom{0} \pm 0.002\phantom{0} \cdot 10^{-2\phantom{0}} $ & $\rm 9.130\phantom{0} \pm 0.004\phantom{0} \cdot 10^{-2\phantom{0}} $\\ 
      \multirow{2}{*}{Cu}
      & prompt & $\rm 7.3\phantom{000} \pm 0.1\phantom{000} \cdot 10^{-1\phantom{0}} $ & $\rm 1.23\phantom{00} \pm 0.13\phantom{00} \cdot 10^{0\phantom{-0}} $ & $\rm 2.20\phantom{00} \pm 0.17\phantom{00} \cdot 10^{0\phantom{-0}} $ & $\rm 2.44\phantom{00} \pm 0.19\phantom{00} \cdot 10^{0\phantom{-0}} $ & $\rm 4.88\phantom{00} \pm 0.29\phantom{00} \cdot 10^{0\phantom{-0}} $ & $\rm 6.09\phantom{00} \pm 0.34\phantom{00} \cdot 10^{0\phantom{-0}} $ & $\rm 1.22\phantom{00} \pm 0.05\phantom{00} \cdot 10^{1\phantom{-0}} $\\
      & decay  & $\rm 6.44\phantom{00} \pm 0.03\phantom{00} \cdot 10^{-3\phantom{0}} $ & $\rm 1.073\phantom{0} \pm 0.005\phantom{0} \cdot 10^{-2\phantom{0}} $ & $\rm 1.93\phantom{00} \pm 0.01\phantom{00} \cdot 10^{-2\phantom{0}} $ & $\rm 2.15\phantom{00} \pm 0.01\phantom{00} \cdot 10^{-2\phantom{0}} $ & $\rm 4.29\phantom{00} \pm 0.02\phantom{00} \cdot 10^{-2\phantom{0}} $ & $\rm 5.366\phantom{0} \pm 0.026\phantom{0} \cdot 10^{-2\phantom{0}} $ & $\rm  1.073\phantom{0} \pm 0.005\phantom{0} \cdot 10^{-1\phantom{0}} $\\
      \multirow{2}{*}{Fe}
      & prompt & $\rm 1.69\phantom{00} \pm 0.12\phantom{00} \cdot 10^{0\phantom{-0}} $ & $\rm 2.84\phantom{00} \pm 0.16\phantom{00} \cdot 10^{0\phantom{-0}} $ & $\rm 5.1\phantom{000} \pm 0.2\phantom{000} \cdot 10^{0\phantom{-0}} $ & $\rm 5.7\phantom{000} \pm 0.2\phantom{000} \cdot 10^{0\phantom{-0}} $ & $\rm 1.13\phantom{00} \pm 0.03\phantom{00} \cdot 10^{1\phantom{-0}} $ & $\rm  1.412\phantom{0} \pm 0.037\phantom{0} \cdot 10^{1\phantom{-0}} $ & $\rm 2.82\phantom{00} \pm 0.05\phantom{00} \cdot 10^{1\phantom{-0}} $\\
      & decay  & $\rm 2.34\phantom{00} \pm 0.06\phantom{00} \cdot 10^{-4\phantom{0}} $ & $\rm 3.9\phantom{000} \pm 0.1\phantom{000} \cdot 10^{-4\phantom{0}} $ & $\rm 7.0\phantom{000} \pm 0.2\phantom{000} \cdot 10^{-4\phantom{0}} $ & $\rm 7.80\phantom{00} \pm 0.21\phantom{00} \cdot 10^{-4\phantom{0}} $ & $\rm 1.56\phantom{00} \pm 0.04\phantom{00} \cdot 10^{-3\phantom{0}} $ & $\rm 1.95\phantom{00} \pm 0.05\phantom{00} \cdot 10^{-3\phantom{0}} $ & $\rm 3.9\phantom{000} \pm 0.1\phantom{000} \cdot 10^{-3\phantom{0}} $\\
      \multirow{2}{*}{Mg}
      & prompt & $\rm 1.61\phantom{00} \pm 0.12\phantom{00} \cdot 10^{0\phantom{-0}} $ & $\rm 2.68\phantom{00} \pm 0.17\phantom{00} \cdot 10^{0\phantom{-0}} $ & $\rm 4.84\phantom{00} \pm 0.23\phantom{00} \cdot 10^{0\phantom{-0}} $ & $\rm 5.38\phantom{00} \pm 0.24\phantom{00} \cdot 10^{0\phantom{-0}} $ & $\rm 1.08\phantom{00} \pm 0.03\phantom{00} \cdot 10^{1\phantom{-0}} $ & $\rm 1.345\phantom{0} \pm 0.038\phantom{0} \cdot 10^{1\phantom{-0}} $ & $\rm 2.67\phantom{00} \pm 0.05\phantom{00} \cdot 10^{1\phantom{-0}} $\\ 
      & decay  & $\rm 3.19\phantom{00} \pm 0.03\phantom{00} \cdot 10^{-2\phantom{0}} $ & $\rm 5.32\phantom{00} \pm 0.05\phantom{00} \cdot 10^{-2\phantom{0}} $ & $\rm 9.56\phantom{00} \pm 0.09\phantom{00} \cdot 10^{-2\phantom{0}} $ & $\rm 1.06\phantom{00} \pm 0.01\phantom{00} \cdot 10^{-1\phantom{0}} $ & $\rm 2.12\phantom{00} \pm 0.02\phantom{00} \cdot 10^{-1\phantom{0}} $ & $\rm 2.652\phantom{0} \pm 0.025\phantom{0} \cdot 10^{-1\phantom{0}} $ & $\rm 5.279\phantom{0} \pm 0.049\phantom{0} \cdot 10^{-1\phantom{0}} $\\    
      \multirow{2}{*}{Mn}
      & prompt & $\rm 1.77\phantom{00} \pm 0.06\phantom{00} \cdot 10^{1\phantom{-0}} $ & $\rm 2.95\phantom{00} \pm 0.08\phantom{00} \cdot 10^{1\phantom{-0}} $ & $\rm 5.30\phantom{00} \pm 0.11\phantom{00} \cdot 10^{1\phantom{-0}} $ & $\rm 5.89\phantom{00} \pm 0.12\phantom{00} \cdot 10^{1\phantom{-0}} $ & $\rm 1.18\phantom{00} \pm 0.02\phantom{00} \cdot 10^{2\phantom{-0}} $ & $\rm 1.48\phantom{00} \pm 0.02\phantom{00} \cdot 10^{2\phantom{-0}} $ & $\rm 2.95\phantom{00} \pm 0.03\phantom{00} \cdot 10^{2\phantom{-0}} $\\ 
      & decay  & $\rm 9.3\phantom{000} \pm 0.1\phantom{000} \cdot 10^{0\phantom{-0}} $ & $\rm 1.56\phantom{00} \pm 0.02\phantom{00} \cdot 10^{1\phantom{-0}} $ & $\rm 2.80\phantom{00} \pm 0.03\phantom{00} \cdot 10^{1\phantom{-0}} $ & $\rm 3.114\phantom{0} \pm 0.036\phantom{0} \cdot 10^{1\phantom{-0}} $ & $\rm 6.23\phantom{00} \pm 0.07\phantom{00} \cdot 10^{1\phantom{-0}} $ & $\rm 7.79\phantom{00} \pm 0.09\phantom{00} \cdot 10^{1\phantom{-0}} $ & $\rm 1.56\phantom{00} \pm 0.02\phantom{00} \cdot 10^{2\phantom{-0}} $\\   
      \multirow{2}{*}{Si}
      & prompt & $\rm 2.75\phantom{00} \pm 0.18\phantom{00} \cdot 10^{-1\phantom{0}} $ & $\rm 4.52\phantom{00} \pm 0.23\phantom{00} \cdot 10^{-1\phantom{0}} $ & $\rm 8.1\phantom{000} \pm 0.3\phantom{000} \cdot 10^{-1\phantom{0}} $ & $\rm 9.1\phantom{000} \pm 0.3\phantom{000} \cdot 10^{-1\phantom{0}} $ & $\rm 1.815\phantom{0} \pm 0.046\phantom{0} \cdot 10^{0\phantom{-0}} $ & $\rm 2.27\phantom{00} \pm 0.05\phantom{00} \cdot 10^{0\phantom{-0}} $ & $\rm 4.55\phantom{00} \pm 0.07\phantom{00} \cdot 10^{0\phantom{-0}} $\\
      & decay  & $\rm 1.6812 \pm 0.0007 \cdot 10^{-6\phantom{0}} $ & $\rm 2.802\phantom{0} \pm 0.001\phantom{0} \cdot 10^{-6\phantom{0}} $ & $\rm 5.038\phantom{0} \pm 0.002\phantom{0} \cdot 10^{-6\phantom{0}} $ & $\rm 5.604\phantom{0} \pm 0.002\phantom{0} \cdot 10^{-6\phantom{0}} $ & $\rm 1.1207 \pm 0.0004 \cdot 10^{-5\phantom{0}} $ & $\rm 1.4008 \pm 0.0006 \cdot 10^{-5\phantom{0}} $ & $\rm 2.801\phantom{0} \pm 0.001\phantom{0} \cdot 10^{-5\phantom{0}} $\\    
      \multirow{2}{*}{Ti}
      & prompt & $\rm 2.60\phantom{00} \pm 0.15\phantom{00} \cdot 10^{0\phantom{-0}} $ & $\rm 4.4\phantom{000} \pm 0.2\phantom{000} \cdot 10^{0\phantom{-0}} $ & $\rm 7.8\phantom{000} \pm 0.3\phantom{000} \cdot 10^{0\phantom{-0}} $ & $\rm 8.70\phantom{00} \pm 0.35\phantom{00} \cdot 10^{0\phantom{-0}} $ & $\rm 1.75\phantom{00} \pm 0.05\phantom{00} \cdot 10^{1\phantom{-0}} $ & $\rm 2.18\phantom{00} \pm 0.06\phantom{00} \cdot 10^{1\phantom{-0}} $ & $\rm 4.36\phantom{00} \pm 0.09\phantom{00} \cdot 10^{1\phantom{-0}} $\\
      & decay  & $\rm 1.595\phantom{0} \pm 0.008\phantom{0} \cdot 10^{-3\phantom{0}} $ & $\rm 2.66\phantom{00} \pm 0.01\phantom{00} \cdot 10^{-3\phantom{0}} $ & $\rm 4.779\phantom{0} \pm 0.025\phantom{0} \cdot 10^{-3\phantom{0}} $ & $\rm 5.316\phantom{0} \pm 0.028\phantom{0} \cdot 10^{-3\phantom{0}} $ & $\rm 1.063\phantom{0} \pm 0.006\phantom{0} \cdot 10^{-2\phantom{0}} $ & $\rm 1.329\phantom{0} \pm 0.007\phantom{0} \cdot 10^{-2\phantom{0}} $ & $\rm 2.66\phantom{00} \pm 0.01\phantom{00} \cdot 10^{-2\phantom{0}} $\\ 
      \multirow{2}{*}{Zn}
      & prompt & $\rm 4.93\phantom{00} \pm 1.38\phantom{00} \cdot 10^{-1\phantom{0}} $ & $\rm 8.3\phantom{000} \pm 1.9\phantom{000} \cdot 10^{-1\phantom{0}} $ & $\rm 1.49\phantom{00} \pm 0.27\phantom{00} \cdot 10^{0\phantom{-0}} $ & $\rm 1.66\phantom{00} \pm 0.29\phantom{00} \cdot 10^{0\phantom{-0}} $ & $\rm 3.32\phantom{00} \pm 0.43\phantom{00} \cdot 10^{0\phantom{-0}} $ & $\rm 4.13\phantom{00} \pm 0.48\phantom{00} \cdot 10^{0\phantom{-0}} $ & $\rm 8.3\phantom{000} \pm 0.7\phantom{000} \cdot 10^{0\phantom{-0}} $\\
      & decay  & $\rm 1.114\phantom{0} \pm 0.008\phantom{0} \cdot 10^{-3\phantom{0}} $ & $\rm 1.86\phantom{00} \pm 0.01\phantom{00} \cdot 10^{-3\phantom{0}} $ & $\rm 3.338\phantom{0} \pm 0.025\phantom{0} \cdot 10^{-3\phantom{0}} $ & $\rm 3.71\phantom{00} \pm 0.03\phantom{00} \cdot 10^{-3\phantom{0}} $ & $\rm 7.42\phantom{00} \pm 0.06\phantom{00} \cdot 10^{-3\phantom{0}} $ & $\rm 9.28\phantom{00} \pm 0.07\phantom{00} \cdot 10^{-3\phantom{0}} $ & $\rm 1.86\phantom{00} \pm 0.01\phantom{00} \cdot 10^{-2\phantom{0}} $
    
    \end{tabular}}
\end{sidewaystable}

Figure~\ref{arco2_p}~and~Table~\ref{tab:yield} %it is also observable, that as
demonstrate that, as both the prompt and the decay gamma yield are determined by the neutron absorption cross section, their energy dependence follows the 1/v rule within the observed energy range in case of all the isotopes of the Ar/CO$_2$ counting gas. Therefore activation with cold neutrons produces a higher yield, and the thermal fraction is negligible.
%%\FloatBarrier
\begin{table}[htbp]
  \centering
  \caption{Major endpoint energies and reaction energies of the main beta-emitters in Ar/CO$_2$ and in aluminium alloy Al5754~\cite{toi}.}
  \label{tab:beta}
  \begin{tabular}{ccccccc}
    \hline
    Isotope  & Reaction  & $\rm Q_{\beta}$         & $\rm E_{\beta}$ &   abundance & $\rm E_{\gamma}$ &  abundance  \\
             & product   &   [keV]                &      [keV]     &             &      [keV]      &              \\
    \hline
    $^{40}$Ar & $^{41}$Ar & $\rm 2491.6 \pm 0.7$   &       1197     &    99~\%    &       1293      &  99~\%       \\
             &           &                        &        814     &  0.0525~\%  &       1677      & 0.0525~\%    \\
             &           &                        &       2491     &  0.8~\%     &        -        &    -         \\
    $^{14}$C  & $^{15}$C  & $\rm 9771.7 \pm 0.8$   &       4472.88  &  63.2~\%    &       5297.817  &  63.2~\%     \\
             &           &                        &       9771.7   &  36.8~\%    &        -        &    -         \\
    $^{18}$O  & $^{19}$O  & $\rm 4821 \pm 3$       &       3266.96  &  54~\%      &       1356.9    &  50~\%       \\
             &           &                        &                &             &        197.1    &  96~\%       \\
             &           &                        &       4623.86  &  45~\%      &        197.1    &  96~\%       \\
    \hline
    $^{27}$Al & $^{28}$Al & $\rm 4642.24 \pm 0.14$ &       2863.21  & 100~\%      &       1778.969  & 100~\%       \\
    $^{55}$Mn & $^{56}$Mn & $\rm 3695.5 \pm 0.3$   &        735.58  &  14.6~\%    &       2113.123  &  14.3~\%     \\
             &           &                        &                &             &        846.771  &  98.9~\%     \\
             &           &                        &       1037.94  &  27.9~\%    &       1810.772  &  27.2~\%     \\
             &           &                        &                &             &        846.771  &  98.9~\%     \\
             &           &                        &       2848.72  &  56.3~\%    &        846.771  &  98.9~\%      
    
  \end{tabular}
\end{table}

As it has been indicated, most of the activated nuclei are beta emitters, and some of the isotopes in the Ar/CO$_{2}$ are pure beta emitters, therefore the effect of beta radiation should also be %observed.
evaluated. In Table~\ref{tab:beta}, the activated beta-emitter isotopes in Ar/CO$_{2}$ and the most significant ones of them in aluminium housing have been collected. As an example, according to the calculated activity concentrations (see Figure~\ref{arco2_act}), only $^{41}$Ar has a considerable activity in the counting gas. Therefore the only beta that might be taken into account is the 1197~keV $^{41}$Ar beta. However, with the usual threshold settings~\cite{piscitelli2013} of proportional systems, the energy-deposition of the beta-radiation does not appear in the measured signal. Therefore on the one hand, the effect of beta radiation is negligible in terms of the detector signal-to-background ratio, while on the other hand, in terms of radiation protection, due to the few 10~cm absorption length in gas and few millimeters absorption length in aluminium, the beta exposure from the detector is also negligible.

\begin{figure}[ht!]
\centering
\includegraphics[width=\textwidth]{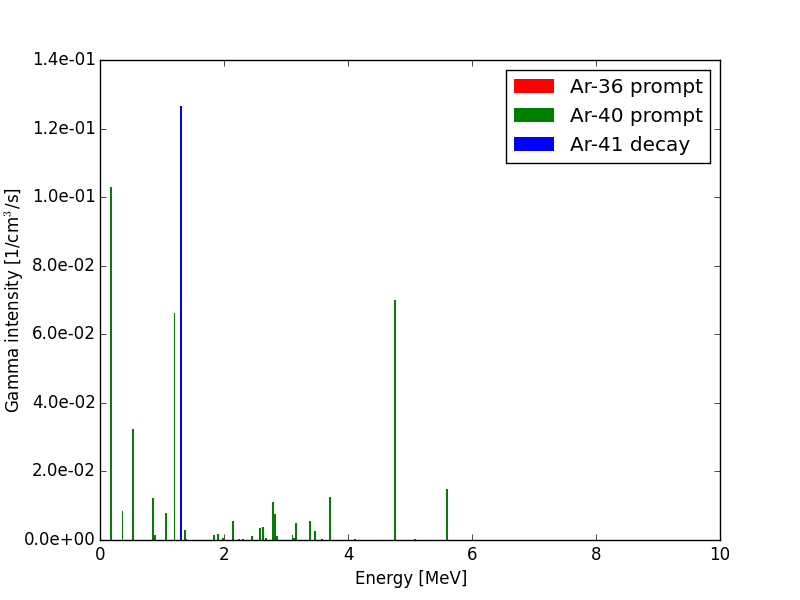}
\caption{Overall prompt and saturated decay gamma spectrum from natural argon, irradiated with $\rm 10^{4}~n/cm^{2}/s$ flux of 1.8~$\rm \AA$ neutrons. Result of calculation on the basis of reaction rates, simulated with MCNP6.1 and decay constant data from Table of Isotopes~\cite{toi}, as explained in text. (For interpretation of the references to color in this figure caption, the reader is referred to the web version of this article.) \label{ar_p_d}}
\end{figure}

Consequently only the prompt and the decay gamma emission have considerable yield to the measured background spectrum, and both of them are dominated by the $^{41}$Ar, during and after the irradiation. A typical neutron beam-on gamma emission spectrum is shown in Figure~\ref{ar_p_d}, for 1.8~$\rm \AA$, $\rm 10^{4} \ n/cm^2/s$ incident neutron flux, calculated with saturated $^{41}$Ar activity.

In order to demonstrate how the gamma radiation background, induced by neutrons in the detector itself, affects the measured signal, signal-to-background ratio has been calculated for detector-filling gas, on the basis of Equations~\ref{eq:S_n}~and~\ref{eq:S_g}. As afore described, Ar/CO$_{2}$ can be represented with $^{41}$Ar in terms of gamma emission. According to its very small saturation time, both the prompt and the decay gamma production have been considered in the background. 
\FloatBarrier
\begin{figure}[ht!]
\centering
\includegraphics[width=\textwidth]{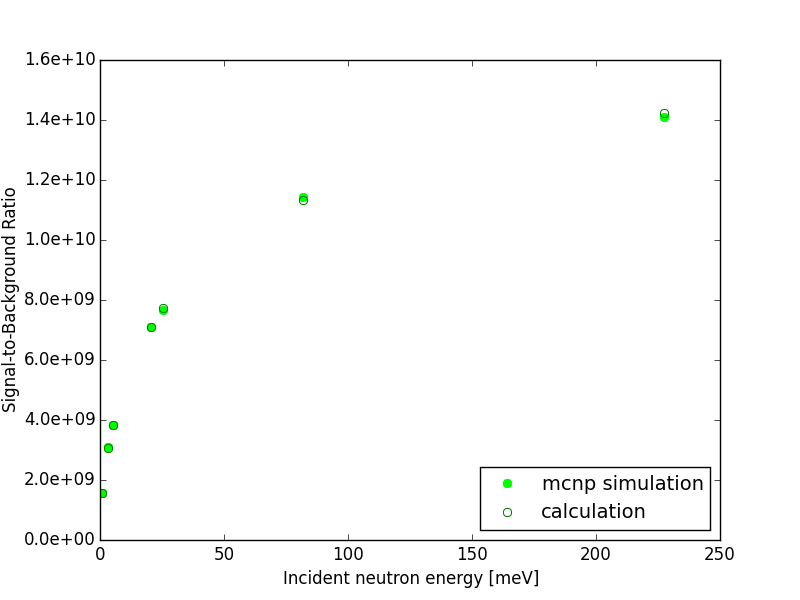}
\caption{Simulated and calculated self-induced signal-to-background ratio for total gamma emission in argon, irradiated with $\rm 10^{4}~n/cm^{2}/s$ neutron flux. (For interpretation of the references to color in this figure caption, the reader is referred to the web version of this article.) \label{SBR_tot}}
\end{figure}

%%Revision
%%In Figure~\ref{SBR_tot} it is shown, that the calculated and the simulated the signal-to-background ratios of the self-induced signal due to activation %are giving good agreement.
%%agree well.
In Figure~\ref{SBR_tot} the good agreement of the calculated and the simulated signal-to-background ratios are shown, for the self-induced gamma background coming from neutron activation. For both cases, %%End of Revised
the signal-to-background ratio increases with the square root of the energy and %changes within the range of
varies between $\rm 10^{9}-10^{10}$ through the entire energy range. %The character of the energy dependence is given by the fact, that the rate of photon production determined by the absorbtion cross section, therefore goes with $\rm 1/ \sqrt{E}$, while for this very specific geometry of the single grid, the energy-dependence of the neutron efficiency is practically linear in the given energy range. %%Althoug the determined signal-to-background ratio is very beneficial, it is less good in the cold neutron, therefore the further developoement of signal-to-background ratio should be considered.
The calculation has been done with a $\rm 10^{-1}$ order of magnitude neutron efficiency, that is typical for a well-designed boron-carbide based neutron detector, and it has been shown that the effect of gamma background is really small, giving only a negligible contribution to the measured signal. Moreover, applying the same calculation for beam monitors, having the lowest possible neutron efficiency (approximated as $\rm 10^{-5}$), the signal-to-background ratio is still $\rm 10^{5}$, meaning that even for beam monitors the self-induced gamma background is vanishingly small.

%% REVISED!
%%\paragraph{Al5754 aluminium frame}
\subsection{Prompt gamma intensity in Al5754 aluminium frame}

%%In order to be able to compare the prompt and decay photon yield of the Ar/CO$_2$ counting gas to the yield of the aluminium frame or housing of the detectors, analytical calculation and MCNP6.1 simluation has been done with the same methods and parameters as the ones used for the Ar/CO$_2$.
The prompt and decay photon yield of the aluminium frame or housing of the detectors have been determined via analytical calculation and MCNP6.1 simulation with the same methods and parameters as the ones used for the Ar/CO$_2$.
%%2016.11.17.
%%The target of irradiation was an aluminium box with 5~mm wall thickness, surrounding the Ar/CO$_2$ gas volume (see Figure~\ref{geom_al}). Alloy Al5754 has been chosen as a typical alloy used in nuclear science.  
Prompt photon production normalised with incident neutron flux has been calculated.
%%Figures~\ref{al5754_P1_calc}-\ref{al5754_P1_mcnp} show the produced prompt photon spectra for $\rm \Phi = 1~n/cm^2/s$ neutron flux, irradiating a 1~cm$^3$ volume, as it is given for Ar/CO$_{2}$ in Figures~\ref{Ar_P_calc}-\ref{O_P_mcnp}. Figures~\ref{al5754_P1_calc}-\ref{al5754_P1_mcnp} are given as an example to show, that qualitatively the calculated and MCNP6.1 simulated spectra agree, although the agreement within the total prompt photon production vary from element to element, as it is shown in Table~\ref{tab:diffph}. Even with the best fitting choice of cross section databases (Table~\ref{tab:crosseclib_al}), for most elements the difference is not higher than 10~\%, but for Mn and Zn the differences between the prompt photon productions are 28~\% and 23~\%, respectively. However, since for all isotopes of these elements the simulation results are conservative, the MCNP simulation remains reliable.

\begin{figure}[ht!]
  \centering
  \includegraphics[width=\textwidth]{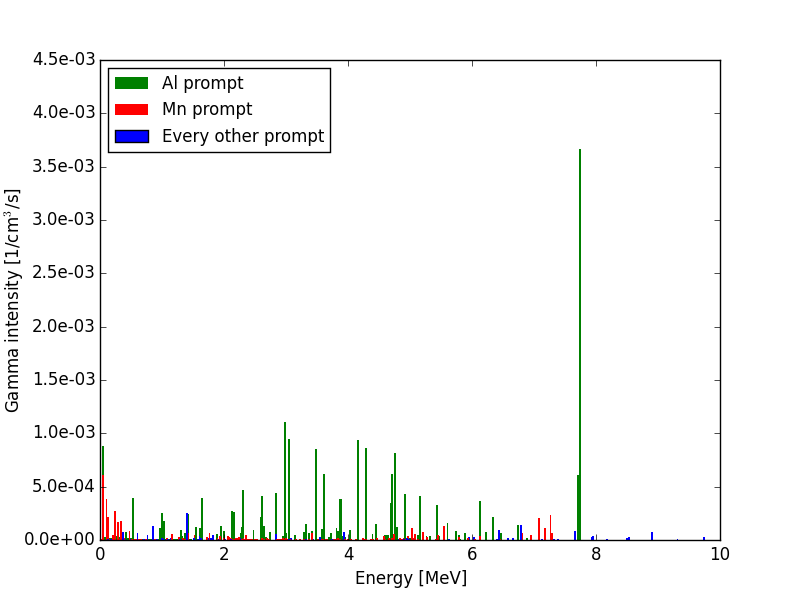}
  \caption{Prompt photon emission spectra from Al5754 aluminium alloy, irradiated with unit flux of 1.8~$\rm \AA$ neutrons. Results of MCNP6.1 simulation, as explained in text. (For interpretation of the references to color in this figure caption, the reader is referred to the web version of this article.) \label{al5754_P1_mcnp} }
\end{figure}

For the Al5754 alloy as well, the calculated and MCNP6.1 simulated spectra qualitatively agree, although the agreement within the total prompt photon production varies from element to element, as shown in Table~\ref{tab:diffph}. Even with the best fitting choice of cross section databases (Table~\ref{tab:crosseclib_al}), the difference is not higher than 10~\% for most elements, but for Mn and Zn the differences between the prompt photon productions are 28~\% and 23~\%, respectively. However, since for all isotopes of these elements the simulation results are conservative, the MCNP simulation remains reliable. Figure~\ref{al5754_P1_mcnp} is given as an example to show the produced prompt photon spectrum for $\rm \Phi = 1~n/cm^2/s$ neutron flux, irradiating an 1~cm$^3$ volume.
\FloatBarrier

Comparing the prompt photon emission from a unit volume of Al5754 %%(see Figures~\ref{al5754_P1_calc}-\ref{al5754_P1_mcnp})
with the same for Ar/CO$_{2}$ %%(Figures~\ref{Ar_P_calc}-\ref{O_P_mcnp}),
(see Table~\ref{tab:yield}) it can be stated, that the prompt photon intensity coming from the aluminium housing is 3~orders of magnitude higher than the one coming from the counting gas. However, for large area detectors, like the ones used in chopper spectrometry, where the gas volume might be $\rm 10^{5}~cm^{3}$ (see~\cite{VORprop,VOR2015,CSPECprop,TREXprop}) the prompt photon yield of the detector counting  gas can become comparable to %the one
that of the solid frame.

\begin{table}[htbp]
  \centering
  \caption{Elemental composition of Al5754 \cite{al5754}, where m\% is the mass fraction of each element in the alloy, and $\rm \Delta I_{ph}$ is the maximum difference between calculated and simulated (MCNP6.1) total prompt photon production for all elements.}
  \label{tab:diffph}
  \begin{tabular}{ccr}
    \hline
    Element & m\% & $\rm \Delta I_{ph}$ \\
    \hline
    Al & 97.4  &  10~\% \\
    Cr &  0.3  &  5~\% \\
    Cu &  0.1  &  9~\% \\
    Fe &  0.4  &  5~\% \\
    Mg &  3.6  &  6~\% \\
    Mn &  0.5  & 28~\% \\
    Si &  0.4  & 10~\% \\
    Ti &  0.15 & 10~\% \\
    Zn &  0.2  & 23~\% 
  \end{tabular}
\end{table}

\begin{figure}[ht!]
\centering
\includegraphics[width=\textwidth]{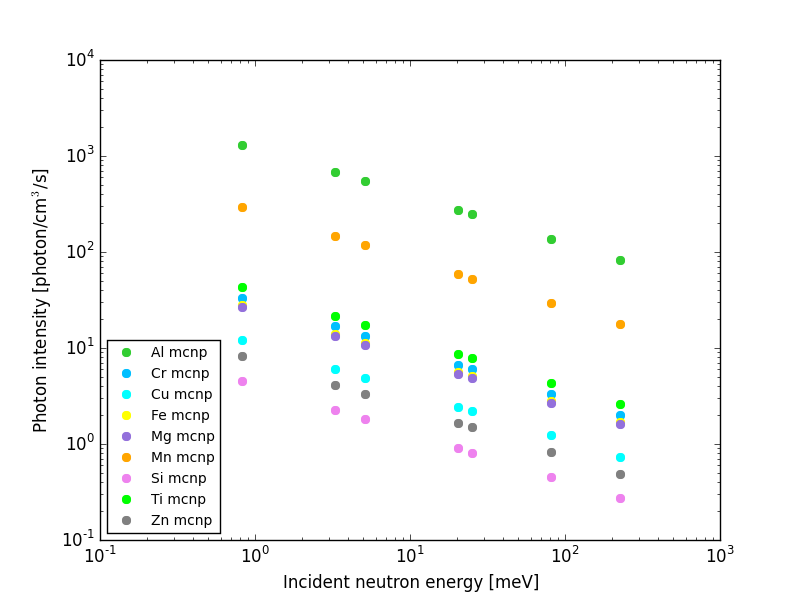}
%%Revised \caption{Elemental distribution of total prompt photon intensity in Al5754 aluminium alloy, irradiated with $\rm 10^{4} n/cm^{2}/s$ flux of neutrons. Results of MCNP6.1 simulation and analytical calculations based on IAEA PGAA Database, as explained in text \cite{iaeapgaa}. \label{al5754_p} }
\caption{Elemental distribution of total prompt photon intensity in Al5754 aluminium alloy, irradiated with $\rm 10^{4}~n/cm^{2}/s$ flux of neutrons. Results of MCNP6.1 simulation, as explained in text. (For interpretation of the references to color in this figure caption, the reader is referred to the web version of this article.) \label{al5754_p} }
\end{figure}

%In Figure~\ref{al5754_p} it is also shown, that
The two main contributors to the prompt photon emission are the aluminium and the manganese (Figure~\ref{al5754_p}); the aluminium total prompt photon yield is 2~order of magnitudes, while the manganese total prompt photon yield is 1~order of magnitude higher than the yield of the rest,
respectively.
Consequently, even the minor components in the aluminium alloy can be relevant for photon production,
if they are having a considerable neutron capture cross section.
%According to Figure~\ref{al5754_P1_mcnp}, within the Al5754 prompt gamma spectrum, there are two main gamma lines that are responsible for the majority of the emission, both belong to $^{27}$Al; the ones with 30.638~$\pm$~0.001~keV and 7724.03~$\pm$~0.04~keV energies. It has to be mentioned, that altough the lower main gamma line appear both in the simulated and the calculated spectra, it only has a significant yield on the basis of IAEA Data, that is not reproduced within the simulation. However, the mentioned gamma energy is low enough that for practical purposes the MCNP simulation remains reliable.
According to Figure~\ref{al5754_P1_mcnp}, within the simulated Al5754 prompt gamma spectrum, there is one main gamma line that is responsible for the majority of the emission, 7724.03~$\pm$~0.04~keV line of $^{27}$Al. 
%the ones with 30.638~$\pm$~0.001~keV and  energies.
It has to be mentioned, that in the analytically calculated spectrum a second main gamma line appears at 30.638~$\pm$~0.001~keV, also from $^{27}$Al; it only has a significant yield on the basis of IAEA Data, that is not reproduced within the simulation. However, the mentioned gamma energy is low enough that for practical purposes the MCNP simulation remains reliable.
\FloatBarrier

%% activity + decay
\subsection{Activity concentration and decay gammas in Al5754 aluminium frame}
An analytical calculation has been performed in order to determine the induced activity in the irradiated aluminium housing, as well as the photon yield coming from the activated radionuclides, with the same methods that have been used for the counting gas. The calculation was based on the bibliographical thermal neutron capture cross sections and the half-lives of the isotopes in the AL5754 aluminium alloy (see Table~\ref{tab:siglambda_al}).
%%2016.11.17
%%Total and isotopical activity concentration have been calculated for a $\rm 10^6~s$ irradiation time (11~days), followed by a $\rm 10^7$~s (0.3~year) cooling time, with $\rm 10^4~n/cm^2/s$ monoenergetic irradiating neutron flux at 0.6, 1, 1.8, 2, 4, 5 and 10~$\rm \AA$ wavelengths.

\begin{figure}[ht!]
\centering
\includegraphics[width=\textwidth]{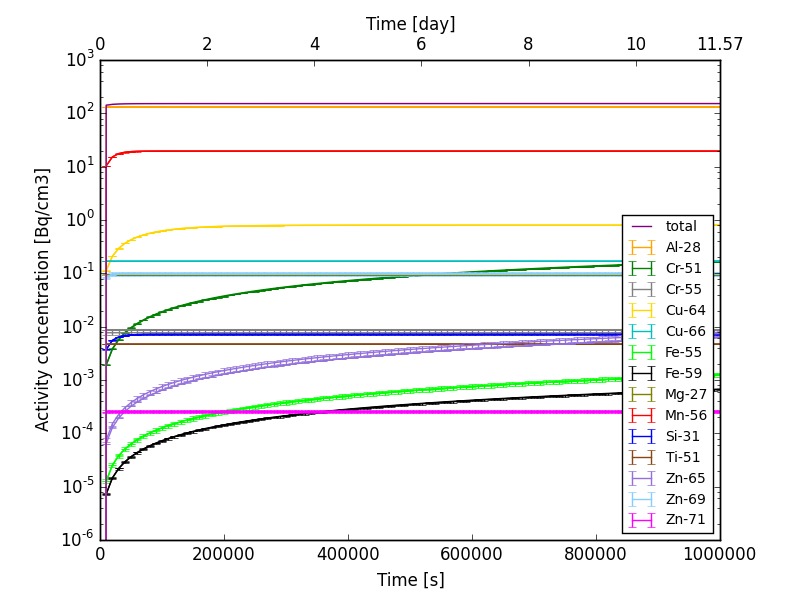}
%%\caption{Build up of total activity concentration in Al5754 during irradiation time. Results of MCNP simulation and analytical calculations \cite{mugh81} \cite{toi}. \label{al5754_act}}
\caption{Build-up of isotopic and total activity concentration [Bq/cm$^3$] in Al5754 aluminium alloy during a 10$^6$~s irradiation time. Results of MCNP6.1 simulation and analytical calculations~\cite{mugh81,toi}, as explained in text. (For interpretation of the references to color in this figure caption, the reader is referred to the web version of this article.) \label{al5754_act} }
\end{figure}

\begin{figure}[ht!]
\centering
\includegraphics[width=\textwidth]{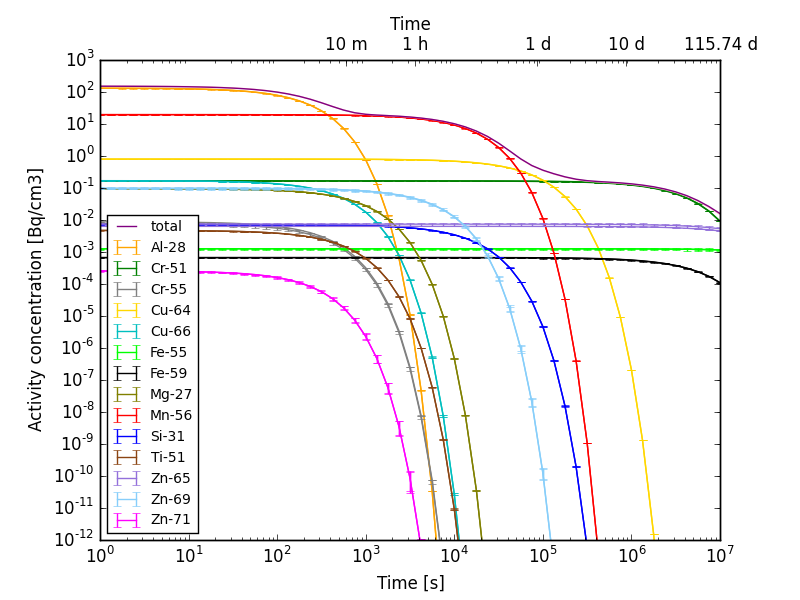}
%%\caption{Decrease of activity concentration in irradiated Al5754 from end of irradiation. Results of MCNP simulation and analytical calculations \cite{mugh81}. \label{al5754_decay}}
\caption{Decrease of activity concentration [Bq/cm$^3$] in Al5754 aluminium alloy from end of the 10$^6$~s irradiation period. Results of MCNP6.1 simulation and analytical calculations~\cite{mugh81,toi}, as explained in text. (For interpretation of the references to color in this figure caption, the reader is referred to the web version of this article.) \label{al5754_decay}}
\end{figure}

An example of the activity build-up during irradiation time for 1.8~$\rm \AA$ is presented in Figure~\ref{al5754_act} for all the produced radionuclei. According to Figures~\ref{al5754_act} and \ref{al5754_decay}, for most of the isotopes in Al5754 the activity concentrations obtained from calculations and MCNP6.1 simulations agree within the margin of error or within the range of 5~\%. However, for a few  isotopes the difference is significant. In the case of $^{55}$Cr with the most suitable choice of cross section libraries largest discrepancy between the simulations and the calculations~\cite{mugh81} is 13~\%. %Zinc is the most problematic element of the simulation;
Also extra care is needed when treating Zn in the simulations; with calculations made on the basis of the thermal neutron cross section data of Mughabghab~\cite{mugh81}, the discrepancies for $^{65}$Zn, $^{69}$Zn, $^{71}$Zn are 5~\%, 7~\% and 10~\% respectively, while in the case of using the NIST database~\cite{nistsig} for the calculations, the differences were 18~\%, 3~\% and 1~\%. Since $^{64}$Zn, the parent isotope of $^{65}$Zn is the major component in the natural zinc, the usage of the first database is recommended. According to Table~\ref{tab:yield}, the activity concentration of the zinc is 5~orders of magnitude smaller than the highest occurring activity concentration, hence the large difference between the calculated and the simulated result does not have a significant impact on the results of the whole alloy.
\FloatBarrier

In Figure~\ref{al5754_act} it is demonstrated, that the majority of the produced total activity is estimated to be given by the $^{28}$Al and the $^{55}$Mn, $\rm 1.33 \cdot 10^{2} \ Bq/cm^3$ and $\rm 1.96 \cdot 10^{1} \ Bq/cm^3$ at the end of the irradiation time, respectively. It is also shown, that for all isotopes the activity concentration saturates quickly at the beginning of the irradiation time, therefore the decay gamma radiation is also produced practically during the entire irradiation time, with a yield constant in time.

%%In Figure~\ref{Al_D}
The decay gamma intensity of the activated radionuclei from a unit volume has also been calculated, with the activity reached by the end of the irradiation time, like in case of Ar/CO$_{2}$ (see Table~\ref{tab:yield}). It is shown that the decay gamma given by the  $^{28}$Al and $^{55}$Mn; their decay photon emission is 3 and 2~orders of magnitude higher then the rest. The decay gamma spectrum is dominated by the $\rm 1778.969 \pm 0.012$~keV line of $^{28}$Al.

Figure~\ref{al5754_p} and Table~\ref{tab:yield} demonstrate, that for aluminium and manganese the prompt photon production ($\rm 2.47 \cdot 10^{2}$ and $\rm 5.27 \cdot 10^{1} \ \frac{photon/cm^3/s}{n/cm^2/s}$) and the saturated decay gamma production ($\rm 1.33 \cdot 10^{2}$ and $\rm 2.8 \cdot 10^{1} \ \frac{photon/cm^3/s}{n/cm^2/s}$) are comparable; the yield of decay photon is 53-54~\% of that of the prompt photon one, whereas for all the other isotopes the decay gamma production is less than 1~\% compared to the prompt gamma production.

\begin{figure}[ht!]
\centering
\includegraphics[width=\textwidth]{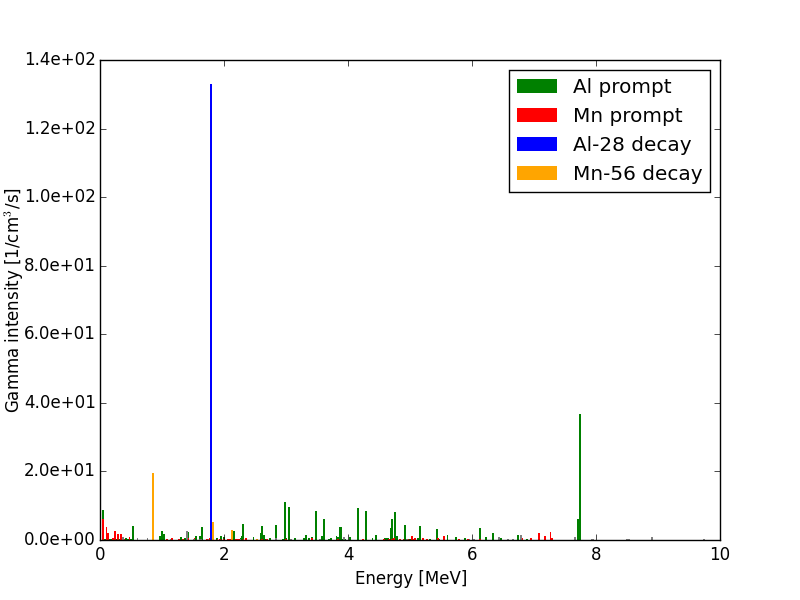}
%%\caption{Prompt and decay gamma spectrum from Al5754. \label{al5754_all_gamma_mcnp}}
\caption{Overall prompt and saturated decay gamma spectrum from Al5754 aluminium alloy, irradiated with $\rm 10^{4}~n/cm^{2}/s$ flux of 1.8~$\rm \AA$ neutrons. Result of calculation on the basis of reaction rates, simulated with MCNP6.1 and decay constant data from Table of Isotopes~\cite{toi}, as explained in text. (For interpretation of the references to color in this figure caption, the reader is referred to the web version of this article.) \label{al5754_all_gamma_mcnp} }
\end{figure}

Figure~\ref{al5754_all_gamma_mcnp} depicts that the total gamma emission spectrum during the neutron irradiation is dominated by the aluminium. The majority of the total photon yield comes from the $^{27}$Al prompt gamma emission, while the two main lines of the measured spectrum are the $\rm 1778.969 \pm 0.012$~keV  $^{28}$Al decay gamma and the 7724.03~$\pm$~0.04~keV $^{27}$Al prompt gamma line.

The decrease of activity in the detector counting gas because of the natural radioactive decay has also been calculated and the obtained results are  shown in Figure~\ref{al5754_decay}, like in the case of Ar/CO$_{2}$ in Figure~\ref{arco2_decay}. There are three isotopes that become major components of the total activity for some period during the cooling time: $^{28}$Al with 1~order of magnitude higher activity than the rest within 0-6$ \cdot 10^3$~s (10~min), $^{56}$Mn with 2~orders of magnitude higher activity than the rest within 6$ \cdot 10^3$-10$^6$~s (11~days), and $^{51}$Cr with 1~order of magnitude higher activity than the rest from 10$^6$~s, therefore the total activity decrease is relatively fast. However, because of the long half-life of $^{55}$Fe, ($\rm T_{\frac{1}{2}} = 2.73 \pm 0.03$~y), a small backround activity is expected to remain for years after the irradiation.

\section{Conclusions}

Analytical calculations and MCNP6.1 modelling have been prepared and compared in order to study the effect of neutron activation in boron-carbide-based neutron detectors. A set of MCNP6.1 cross section databases has been collected for Ar/CO$_{2}$ counting gas and aluminium detector housing estimated as Al5754, which both give good agreement with the analytical calculations, or give an acceptable, conservative estimation both for prompt gamma production and activity calculations. These databases are recommended to use for more complex geometries, where the analytical calculations should be replaced by MCNP simulations.

It has been shown, that the prompt photon emission of the aluminium housing is dominated by the Al and Mn contributors, while that of the counting gas is mainly given by Ar. The prompt photon intensity from an aluminium-housing unit volume is 3~orders of magnitude higher than from that of the counting gas.

 The total activity concentration of the housing is mainly given by the $^{28}$Al and the $^{56}$Mn, and given by the  $^{41}$Ar in the counting gas. Due to the short half-lives of the main isotopes, their decay gammas already appear and saturate during the irradiation period, giving a comparable decay gamma emission to the prompt photon emission in terms of yield.

With the afore mentioned typical counting gas, the decay gamma yield of $^{41}$Ar saturates at $\rm 1.28 \cdot 10^{-1} Bq/cm^{3}$, and based on this value, operational scenarios can be envisaged. With these results it has been shown, that only a low level of activation is expected in the detector counting gas. Therefore with a flushing of 1 detector volume of gas per day, assuming a $\rm V = 10^{7} \ cm^{3}$ detector volume, $\rm 1.28 \cdot 10^{6} \ Bq/day$ activity production is expected. By varying the flush rate and storing the counting gas up to 1~day before release, only negligible levels of activity will be present in the waste Ar/CO$_{2}$ stream.

%Based on the above mentioned results, the
Neutron-induced gamma signal-to-background ratio has also been determined for several neutron energies, revealing that the signal-to-background ratio changes within the range of $\rm 10^9 - 10^{10}$ for general boron-carbide-based detector geometries, and still being $10^{5}$ even for beam monitors, having the lowest possible efficiency.

The effect of beta-radiation coming from the activated isotopes has also been considered, and it can be stated that the beta-radiation is negligible both in terms of the signal-to-background ratio and in terms of radiation protection.

In this study all simulations and calculations were made for a generic geometry, and a reliable set of data on activity and photon production were given that can be generally applied and scaled for any kind or boron-carbide-based neutron detector, filled with Ar/CO$_2$.

\section*{Acknowledgments}

This work has been supported by the In-Kind collaboration between ESS ERIC and the Centre for Energy Research of the Hungarian Academy of Sciences (MTA EK). %%Special thanks to BrightnESS Grant for the support of Richard Hall-Wilton. %Also have to be thanked Szabolcs Czifrus, for giving professional help with MCNP6.1 simulations.
Richard Hall-Wilton would like to acknowledge the EU Horizon2020 Brightness Grant [676548].

%% \section*{References}

%% \bibliography{../Reference}

\newpage
\section*{Appendix}

\setcounter{table}{0}
\renewcommand{\thetable}{A\arabic{table}}

\begin{table}[htbp]
  \centering
  \caption{(n,$\gamma$) reaction cross sections at 25.30 meV~\cite{mugh81}, reaction products, their decay constants and %corresponding
    respective uncertainties~\cite{toi} ($\sigma\rm(^{14}C)$ is from TENDL-2014 database~\cite{iaeands}).}
  \label{tab:siglambda}
  \begin{tabular}{cccccc}
    \hline
    Isotope & Reaction & $\sigma $ & $\Delta \sigma $ &  $\lambda $ & $\Delta \lambda $ \\
            & product  &   [barn] &      [barn]     &  [s$^{-1}$] &    [s$^{-1}$] \\
    \hline
    $^{36}$Ar & $^{37}$Ar & $\rm 5.20 \cdot 10^{0\phantom{-}}$ & $\rm 5.00 \cdot 10^{-1}$ & $\rm 2.29 \cdot 10^{-7\phantom{0}}$ & $\rm 2.61 \cdot 10^{-10}$ \\
    $^{38}$Ar & $^{39}$Ar & $\rm 8.00 \cdot 10^{-1}$ & $\rm 2.00 \cdot 10^{-1}$ & $\rm 8.17 \cdot 10^{-11}$ & $\rm 9.11 \cdot 10^{-13}$ \\
    $^{40}$Ar & $^{41}$Ar & $\rm 6.60 \cdot 10^{-1}$ & $\rm 1.00 \cdot 10^{-2}$ & $\rm 1.06 \cdot 10^{-4\phantom{0}}$ & $\rm 1.16 \cdot 10^{-7\phantom{0}}$ \\
    $^{16}$O & $^{17}$O & $\rm 1.90 \cdot 10^{-4}$ & $\rm 1.90 \cdot 10^{-5}$ & stable & - \\
    $^{17}$O & $^{18}$O & $\rm 5.38 \cdot 10^{-4}$ & $\rm 6.50 \cdot 10^{-5}$ & stable & - \\
    $^{18}$O & $^{19}$O & $\rm 1.60 \cdot 10^{-4}$ & $\rm 1.00 \cdot 10^{-5}$ & $\rm 2.58 \cdot 10^{-2\phantom{0}}$ & $\rm 7.66 \cdot 10^{-5\phantom{0}}$ \\
    $^{12}$C & $^{13}$C & $\rm 3.53 \cdot 10^{-3}$ & $\rm 2.00 \cdot 10^{-3}$ & stable & - \\
    $^{13}$C & $^{14}$C & $\rm 1.37 \cdot 10^{-3}$ & $\rm 4.00 \cdot 10^{-5}$ & $\rm 3.84 \cdot 10^{-12}$ & $\rm 2.67 \cdot 10^{-14}$ \\ 
    $^{14}$C & $^{15}$C & $\rm 8.11 \cdot 10^{-7}$ & - & $\rm 2.83 \cdot 10^{-1\phantom{0}}$ & $\rm 5.78 \cdot 10^{-4\phantom{0}}$
    
  \end{tabular}
\end{table}
%%\FloatBarrier

\begin{table}[htbp]
  \centering
  \caption{Cross section for Ar/CO$_{2}$ libraries used in MCNP6.1 simulations \cite{mcnplib}.}
  \label{tab:crosseclib}
  \begin{tabular}{ccccc}
    \hline
    Element & \multicolumn{2}{c}{Prompt gamma} & \multicolumn{2}{c}{ (n,$\gamma$) reaction rate} \\
            & \multicolumn{2}{c}{production}   & \multicolumn{2}{c}{calculation}                \\
%%    Element & Prompt gamma &              & (n,$\gamma$) reaction rate  &  \\
%%            & production   &              &   calculation              &   \\
    \hline
    Ar      & 18000.42c    &  LANL             & 18036.80c  & ENDF/B-VII.1 \\
            &              &                   & 18038.80c  & ENDF/B-VII.1 \\
            &              &                   & 18040.80c  & ENDF/B-VII.1 \\
    C       &  6000.80c    & ENDF/B-VII.1      &  6012.00c  & TALYS-2015   \\
            &              &                   &  6013.00c  & TALYS-2015   \\
            &              &                   &  6014.00c  & TALYS-2015   \\
    O       &  8000.80c    & ENDF/B-VII.1      &  8016.00c  & TALYS-2015   \\
            &              &                   &  8017.00c  & TALYS-2015   \\
            &              &                   &  8018.00c  & TALYS-2015 
  \end{tabular}
\end{table}
%%\FloatBarrier

\begin{table}[htbp]
  \centering
  \caption{(n,$\gamma$) reaction cross sections at 25.30 meV~\cite{mugh81}, reaction products, their decay constants and %corresponding
    respective uncertainties~\cite{toi}.}
  \label{tab:siglambda_al}
  \begin{tabular}{cccccc}
    \hline
    Isotope & Reaction & $\sigma $ & $\Delta \sigma $ &  $\lambda $ & $\Delta \lambda $ \\
            & product  &   [barn] &      [barn]     &  [s$^{-1}$] &    [s$^{-1}$] \\
    \hline
    $^{27}$Al & $^{28}$Al & $\rm 2.31 \cdot 10^{-1}$ & $\rm 3.00 \cdot 10^{-3}$ & $\rm 1.35 \cdot 10^{2}$ & $\rm 7.20 \cdot 10^{-2}$ \\
    $^{50}$Cr & $^{51}$Cr & $\rm 1.59 \cdot 10^{1\phantom{-}}$ & $\rm 2.00 \cdot 10^{-1}$ & $\rm 2.39 \cdot 10^{6}$ & $\rm 2.07 \cdot 10^{2\phantom{-}}$ \\
    $^{52}$Cr & $^{53}$Cr & $\rm 7.60 \cdot 10^{-1}$ & $\rm 6.00 \cdot 10^{-2}$ & stable                  & -                       \\
    $^{53}$Cr & $^{54}$Cr & $\rm 1.82 \cdot 10^{1\phantom{-}}$ & $\rm 1.50 \cdot 10^{0\phantom{-}}$ & stable                  & -                       \\
    $^{54}$Cr & $^{55}$Cr & $\rm 3.60 \cdot 10^{-1}$ & $\rm 4.00 \cdot 10^{-2}$ & $\rm 2.10 \cdot 10^{2}$ & $\rm 1.80 \cdot 10^{-1}$ \\
    $^{63}$Cu & $^{64}$Cu & $\rm 4.50 \cdot 10^{0\phantom{-}}$ & $\rm 2.00 \cdot 10^{-2}$ & $\rm 4.57 \cdot 10^{4}$ & $\rm 7.20 \cdot 10^{0\phantom{-}}$ \\
    $^{65}$Cu & $^{66}$Cu & $\rm 2.17 \cdot 10^{0\phantom{-}}$ & $\rm 3.00 \cdot 10^{-2}$ & $\rm 3.07 \cdot 10^{2}$ & $\rm 8.40 \cdot 10^{-1}$ \\
    $^{54}$Fe & $^{55}$Fe & $\rm 2.25 \cdot 10^{0\phantom{-}}$ & $\rm 1.80 \cdot 10^{-1}$ & $\rm 8.61 \cdot 10^{7}$ & $\rm 9.46 \cdot 10^{5\phantom{-}}$ \\
    $^{56}$Fe & $^{57}$Fe & $\rm 2.59 \cdot 10^{0\phantom{-}}$ & $\rm 1.40 \cdot 10^{-1}$ & stable                  & -                       \\
    $^{57}$Fe & $^{58}$Fe & $\rm 2.48 \cdot 10^{0\phantom{-}}$ & $\rm 3.00 \cdot 10^{-1}$ & stable                  & -                       \\
    $^{58}$Fe & $^{59}$Fe & $\rm 1.28 \cdot 10^{0\phantom{-}}$ & $\rm 5.00 \cdot 10^{-2}$ & $\rm 3.85 \cdot 10^{6}$ & $\rm 5.18 \cdot 10^{2\phantom{-}}$ \\
    $^{24}$Mg & $^{25}$Mg & $\rm 5.10 \cdot 10^{-2}$ & $\rm 5.00 \cdot 10^{-3}$ & stable                  & -                       \\
    $^{25}$Mg & $^{26}$Mg & $\rm 1.90 \cdot 10^{-1}$ & $\rm 3.00 \cdot 10^{-2}$ & stable                  & -                       \\
    $^{26}$Mg & $^{27}$Mg & $\rm 3.82 \cdot 10^{-2}$ & $\rm 8.00 \cdot 10^{-4}$ & $\rm 5.67 \cdot 10^{2}$ & $\rm 7.20 \cdot 10^{-1}$ \\
    $^{55}$Mn & $^{56}$Mn & $\rm 1.33 \cdot 10^{1\phantom{-}}$ & $\rm 2.00 \cdot 10^{-1}$ & $\rm 9.28 \cdot 10^{3}$ & $\rm 7.20 \cdot 10^{-1}$ \\
    $^{28}$Si & $^{29}$Si & $\rm 1.77 \cdot 10^{-1}$ & $\rm 5.00 \cdot 10^{-3}$ & stable                  & -                       \\
    $^{29}$Si & $^{30}$Si & $\rm 1.01 \cdot 10^{-1}$ & $\rm 1.40 \cdot 10^{-2}$ & stable                  & -                       \\
    $^{30}$Si & $^{31}$Si & $\rm 1.07 \cdot 10^{-1}$ & $\rm 2.00 \cdot 10^{-3}$ & $\rm 9.44 \cdot 10^{3}$ & $\rm 1.80 \cdot 10^{2\phantom{-}}$ \\
    $^{46}$Ti & $^{47}$Ti & $\rm 5.90 \cdot 10^{-1}$ & $\rm 1.80 \cdot 10^{-1}$ & stable                  & -                       \\
    $^{47}$Ti & $^{48}$Ti & $\rm 1.70 \cdot 10^{0\phantom{-}}$ & $\rm 2.00 \cdot 10^{-1}$ & stable                  & -                       \\
    $^{48}$Ti & $^{49}$Ti & $\rm 7.84 \cdot 10^{0\phantom{-}}$ & $\rm 2.50 \cdot 10^{-1}$ & stable                  & -                       \\
    $^{49}$Ti & $^{50}$Ti & $\rm 2.20 \cdot 10^{0\phantom{-}}$ & $\rm 3.00 \cdot 10^{-1}$ & stable                  & -                       \\
    $^{50}$Ti & $^{51}$Ti & $\rm 1.79 \cdot 10^{-1}$ & $\rm 3.00 \cdot 10^{-1}$ & $\rm 3.46 \cdot 10^{2}$ & $\rm 6.00 \cdot 10^{-1}$ \\
    $^{64}$Zn & $^{65}$Zn & $\rm 7.60 \cdot 10^{-1}$ & $\rm 2.00 \cdot 10^{-2}$ & $\rm 2.11 \cdot 10^{7}$ & $\rm 2.25 \cdot 10^{4\phantom{-}}$ \\
    $^{66}$Zn & $^{67}$Zn & $\rm 8.50 \cdot 10^{-1}$ & $\rm 2.00 \cdot 10^{-1}$ & stable                  & -                       \\
    $^{67}$Zn & $^{68}$Zn & $\rm 6.80 \cdot 10^{0\phantom{-}}$ & $\rm 8.00 \cdot 10^{-1}$ & stable                  & -                       \\
    $^{68}$Zn & $^{69}$Zn & $\rm 1.00 \cdot 10^{1\phantom{-}}$ & $\rm 1.00 \cdot 10^{-1}$ & $\rm 3.38 \cdot 10^{3}$ & $\rm 5.40 \cdot 10^{1\phantom{-}}$ \\
    $^{70}$Zn & $^{71}$Zn & $\rm 8.30 \cdot 10^{-2}$ & $\rm 5.00 \cdot 10^{-3}$ & $\rm 1.47 \cdot 10^{2}$ & $\rm 6.00 \cdot 10^{0\phantom{-}}$ 
    %% $^{}$ & $^{}$ & $\rm  \cdot 10^{}$ & $\rm  \cdot 10^{}$ & $\rm  \cdot 10^{}$ & $\rm  \cdot 10^{}$ \\
  \end{tabular}
\end{table}
%%
%%The cross section libraries used for MCNP simulation are shown in Table~\ref{tab:crosseclib_al}.
%%\FloatBarrier

\begin{table}[htbp]
  \centering
  \caption{Cross section libraries for Al5754  used in MCNP6.1 simulations \cite{mcnplib}.}
  \label{tab:crosseclib_al}
  \begin{tabular}{ccccc}
    \hline
    Element & \multicolumn{2}{c}{Prompt gamma} & \multicolumn{2}{c}{ (n,$\gamma$) reaction rate} \\
            & \multicolumn{2}{c}{production}   & \multicolumn{2}{c}{calculation}                \\
    \hline
    Al      & 13027.66c    & ENDF/B-VI.6       & 13027.80c    & ENDF/B-VII.1 \\
    Cr      & 24050.80c    & ENDF/B-VII.1      & 24050.80c    & ENDF/B-VII.1 \\
            & 24052.80c    & ENDF/B-VII.1      & 24052.80c    & ENDF/B-VII.1 \\
            & 24053.80c    & ENDF/B-VII.1      & 24053.80c    & ENDF/B-VII.1 \\
            & 24054.80c    & ENDF/B-VII.1      & 24054.80c    & ENDF/B-VII.1 \\
    Cu      & 29063.80c    & ENDF/B-VII.1      & 29063.80c    & ENDF/B-VII.1 \\
            & 29065.80c    & ENDF/B-VII.1      & 29065.80c    & ENDF/B-VII.1 \\
    Fe      & 26054.80c    & ENDF/B-VII.1      & 26054.00c    & TALYS-2015   \\
            & 26056.80c    & ENDF/B-VII.1      & 26056.00c    & TALYS-2015   \\
            & 26057.80c    & ENDF/B-VII.1      & 26057.00c    & TALYS-2015   \\
            & 26058.80c    & ENDF/B-VII.1      & 26058.00c    & TALYS-2015   \\
    Mg      & 12000.62c    & ENDF/B-VI.8       & 12024.80c    & ENDF/B-VII.1 \\       
            &              &                   & 12025.80c    & ENDF/B-VII.1 \\
            &              &                   & 12026.80c    & ENDF/B-VII.1 \\
    Mn      & 25055.62c    & ENDF/B-VI.8       & 25055.80c    & ENDF/B-VII.1 \\
    Si      & 14000.60c    & ENDF/B-VI.0       & 14028.80c    & ENDF/B-VII.1 \\
            &              &                   & 14029.80c    & ENDF/B-VII.1 \\
            &              &                   & 14030.80c    & ENDF/B-VII.1 \\
    Ti      & 22000.62c    & ENDF/B-VI.8       & 22046.80c    & ENDF/B-VII.1 \\
            &              &                   & 22047.80c    & ENDF/B-VII.1 \\
            &              &                   & 22048.80c    & ENDF/B-VII.1 \\
            &              &                   & 22049.80c    & ENDF/B-VII.1 \\
            &              &                   & 22050.80c    & ENDF/B-VII.1 \\
    Zn      & 30064.80c    & ENDF/B-VII.1      & 30064.00c    & TALYS-2015   \\ 
            & 30066.80c    & ENDF/B-VII.1      & 30066.00c    & TALYS-2015   \\  
            & 30067.80c    & ENDF/B-VII.1      & 30067.00c    & TALYS-2015   \\  
            & 30068.80c    & ENDF/B-VII.1      & 30068.00c    & TALYS-2015   \\  
            & 30070.80c    & ENDF/B-VII.1      & 30070.00c    & TALYS-2015 
  \end{tabular}
\end{table}

\FloatBarrier
\newpage
\section*{References}

\bibliography{../Reference}

\end{document}